\documentclass[sigconf]{acmart}
\usepackage[ruled,linesnumbered,vlined]{algorithm2e}
\renewcommand\footnotetextcopyrightpermission[1]{} 
\graphicspath{{FiguresISO/}}
\DeclareGraphicsExtensions{.eps}
\usepackage{graphicx}
\usepackage{balance}  

\usepackage{balance}  

\usepackage{epsfig,latexsym,graphics,subfigure}
\usepackage{mathrsfs}
\usepackage[normalem]{ulem}
\usepackage{url}
\usepackage{epsfig}
\usepackage{xspace}
\usepackage{color}
\usepackage{pstricks}
\usepackage{times}
\usepackage{epsfig}
\usepackage{xspace}
\usepackage{subfigure}
\usepackage{float}
\usepackage{latexsym}
\usepackage{url}
\usepackage{color}
\usepackage{multirow}
\usepackage{tikz}
\usepackage{colortbl}
\usepackage{tabularx}
\usepackage[noend]{algpseudocode}
\usepackage{amsthm}
\newtheorem{myDef}{Definition}
\newtheorem{myLem}{Lemma}
\newtheorem{myTheo}{Theorem}

\floatname{algorithm}{Algorithm}

\begin{document}
\title{Influence Maximization in Hypergraphs by Stratified Sampling for Efficient Generation of Reverse Reachable Sets}
\author{
Lingling Zhang, Hong Jiang, Ye Yuan and Guoren Wang}
\affiliation{%
  \institution{Capital Normal University,$\ $ University of Texas at Arlington,$\ $ Beijing Institute of Technology}
}
\email{7089@cnu.edu.cn, hong.jiang@uta.edu, {yuan-ye,wanggr}@bit.edu.cn}
\begin{abstract}
Given a hypergraph, influence maximization (IM) is to discover a seed set containing $k$ vertices
that have the maximal influence. Although the existing vertex-based IM algorithms perform better than the hyperedge-based algorithms
by generating random reverse researchable (RR) sets, they are inefficient because (i) they ignore important structural information
associated with hyperedges and thus obtain inferior results, (ii) the frequently-used sampling methods
for generating RR sets have low efficiency because of a large number of required samplings along with high sampling variances, and
(iii) the vertex-based IM algorithms have large overheads in terms of running time and memory costs.
To overcome these shortcomings, this paper proposes a novel approach, called \emph{HyperIM}. The key idea behind \emph{HyperIM} is to differentiate structural information
of vertices for developing stratified sampling combined with highly-efficient strategies to generate the RR sets.
With theoretical guarantees, \emph{HyperIM} is able to accelerate the influence spread, improve the sampling efficiency,
and cut down the expected running time. To further reduce the running time and memory costs, we optimize \emph{HyperIM} by inferring the bound of the required number of RR sets in conjunction with stratified sampling.
Experimental results on real-world hypergraphs show that \emph{HyperIM} is able to reduce the number of required RR sets and running time by orders of magnitude while
increasing the influence spread by up to $2.73X$ on average, compared to the state-of-the-art IM algorithms.
\end{abstract}
\maketitle
\section{Introduction}
\label{sec:introduction}
In recent years, hypergraphs have been found more appropriate and effective than regular graphs to model the connections
in complicated networks such as social networks, literature networks and molecular networks \cite{zhang2023efficiently, he2021click}.
A hyperedge in a hypergraph is able to represent connections among an arbitrary
number of vertices while an edge in a regular graph simply represents the connection between two vertices.
Influence spread is defined as the number of vertices influenced by a vertex set containing $k$ vertices under a cascade model that expresses the effect of information spreading over networks \cite{tang2018online, guo2020influence}.
Influence maximization (IM) in a hypergraph is to discover a seed set among all the vertex sets with the same size $k$ having the maximum influence spread.
IM has important applications, such as broadcasting opinions, promoting products, and discovering the influential users in complicated networks.

\begin{figure}[hbt!]
\centering
\includegraphics[width=0.31\textwidth]{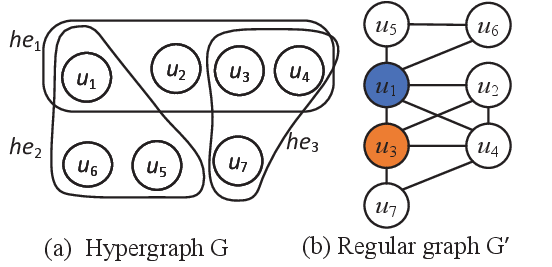}
\caption{Example of a hypergraph, Figure \ref{fig_example}(a) and its transformation into a regular graph,
Figure \ref{fig_example}(b). When analyzing the influence, although the vertices $u_1$ and $u_3$ have the same structure in Hypergraph, as shown in Figure \ref{fig_example}(a), they show different structures shown in the corresponding regular graph in Figure \ref{fig_example}(b).}\label{fig_example}
\end{figure}

Given a hypergraph $G$ with $|V|$ vertices and $|HE|$ hyperedges,
the studies of IM can be classified into two types: \emph{hyperedge-based IM algorithms} and \emph{vertex-based IM algorithms}.
The former tend to estimate the influence from the aspect of hyperedges by discovering the connection among hyperedges \cite{zhu2018social, ma2022hyper,zhang2023influence}
while the latter study the information diffusion process from the aspect of connection among vertices \cite{amato2017influence, ma2022hyper, antelmi2021social}. Since the time complexity of studying influence spreading along the hyperedges is large \cite{aktas2022influence, xie2022influence}, it is impractical for hyperedge-based IM algorithms to discover the seed set for a relatively large hypergraph.
For example, as shown in \cite{xie2022influence}, a representative hyperedge-based IM algorithm costs $15991$ seconds by using a greedy algorithm to discover the seed set in a hypergraph
just containing $1268$ hyperedges. The vertex-based IM algorithms \cite{amato2017influence, ma2022hyper, antelmi2021social}
first describe the connection among vertices by using the corresponding regular graph of a hypergraph such that a hyperedge containing $m\ (m\geq 2)$ vertices can be replaced
by $\frac{m(m-1)}{2}$ regular edges. As shown in Figure \ref{fig_example}, Hypergraph $G$ in Figure \ref{fig_example}(a)
can be transformed into a regular graph $G'$ in Figure \ref{fig_example}(b).
Then, the vertex-based IM algorithms designed for regular graphs, such as \emph{TIM} \cite{tang2015influence} and \emph{IMM} \cite{tang2014influence}, are able to discover the influential vertices from the transformed regular graph.
In contrast to the hyperedge-based IM algorithms that consume huge time but without a desired approximation guarantee,
the vertex-based IM algorithms are popular for discovering the seed set with theoretical guarantees on both the approximation and the time complexity.

Kempe et al. \cite{kempe2003maximizing}, the first study of vertex-based IM algorithms on influence maximization, proves the IM problem to be NP-hard. They employ a general greedy algorithm under two popular cascade models, namely, the Independent-Cascade (IC) model and the Linear- Threshold (LT) model. Thus, the algorithm is able to discover $k$ vertices that have the largest influence under the two models by returning a ($1-\frac{1}{e}-\epsilon$)-approximate solution for any given $\epsilon\in (0,1)$. Their proposed algorithm requires $\Omega(k|V||E| poly(\frac{1}{\epsilon}))$ computation time \cite{kempe2003maximizing} where $|V|$ and $|E|$ denote the numbers of vertices and edges in the regular graph. As the size of the graph increases, the cost of the algorithm becomes extremely high.
Borgs et al. \cite{borgs2014maximizing} make a theoretical breakthrough that reduces the
time complexity of the greedy algorithm to $O(k(|V|+|E|)\log\frac{|V|^{2}}{\epsilon^{3}})$
by generating a sufficiently large number of random reverse reachable (RR) sets, defined as the sets of vertices influencing,
also referred to as activating, some vertices. They still achieve the ($1-\frac{1}{e}-\epsilon$)-approximate solution.
The main step of their proposed algorithm is to use reverse influence sampling to generate RR sets. Since then, many follow-up studies
focus on sampling methods for generating the RR sets. For example, \emph{SUMSIM} \cite{guo2020influence}
proposes a subset sampling method and \emph{TriangleIM} \cite{hu2023triangular} uses a triangle-based sampling method to generate the RR sets. How the RR sets are generated by sampling in a vertex-based IM algorithm is the key factor determining the accuracy of and
time complexity for discovering a seed set.

However, when the existing algorithms are used to discover the seed set in a hypergraph,
they face the following challenges. \emph{First,} these algorithms ignore the important structural information associated with hyperedges,
resulting in inferior vertex selections. As shown in Figure \ref{fig_example}, the vertices
$u_1$ and $u_3$ in hypergraph $G$ are included by two hyperedges. However, when hypergraph $G$ is transformed into the regular graph
$G'$, $u_1$ and $u_3$ have different structures such that $u_1's$ degree is five while $u_3's$ is four. This is because, although $u_4$
is included by two hyperedges, it is counted as one neighbor of $u_3$. Such structural inconsistency makes the existing IM algorithms
inaccurate when selecting the influential vertices. \emph{Second,} the frequently-used sampling methods for generating the RR sets have low sampling efficiency but have high sampling variance.
The probability distribution of the sampling operations in the existing IM algorithms follow either the geometric distribution \cite{gomez2010another} or \emph{(0-1)-}distribution \cite{chen1997statistical}. For example, given a sample set with $n$ vertices,
each of which is sampled with probability $p$, $\lceil\frac{1}{p} \rceil$ sampling operations on average just add one vertex into a RR set while the sampling variance in the geometric-based sampling methods reaches $\frac{np(1-p)}{p^2}$. \emph{Third,} the time complexity is large to generate a large number of required RR sets which in turn results in high memory costs to store these RR sets.

To address these challenges, we propose a novel approach, named \emph{HyperIM}, to efficiently settle the IM problem in hypergraphs. The key idea behind \emph{HyperIM} is to differentiate vertex structures for developing stratified sampling combined with a Binomial-based strategy and a Possion-based strategy that changes how sampling is done to increase sampling efficiency. Specifically, when generating a random RR set of a given vertex, its relevant edges determine the size of the sample set while the number of hyperedges from which each edge is transformed acts as a guide to stratify the vertices in the sample set into different layers. This stratification helps alleviate the impact of post-transformation structural difference and inconsistency mentioned earlier, by separating structurally different vertices of the sample set into different layers. Furthermore, \emph{HyperIM} designs a Binomial-based strategy and a Possion-based strategy to render the probability distributions of the sampling operations on the sample set with $n$ vertices to either follow a binomial distribution ($n\leq 20$) or a Poisson Distribution ($n>20$ ). In contrast to the existing IM algorithms, \emph{HyperIM} optimizes the vertex selections to ensure high influence spread and increases the sampling efficiency while decreasing the sampling variances.

We further show that the time complexity of \emph{HyperIM} is bounded by $O(k\cdot |V|\cdot \theta (l^{*}) \log\frac{|V|}{\epsilon^{2}})$ where $l^{*}$ is a function of the number of layers of the stratified sampling. The time complexity is proven to be smaller than the existing state-of-the-art IM algorithms while providing the same desired theoretical guarantee. Moreover, when dealing with a highly complicated network, even if we use the stratified sampling combined with the two strategies to generate random RR sets,
it still requires substantial computation and memory costs to process the RR sets when the hypergraph is very large. To reduce the number of required RR sets, we first divide the RR sets into two independent parts, explore the stratified information of the influential ability to derive an upper bound and a lower bound of the obtained influence from the divided RR sets, and then compute the maximum number of the RR sets to obtain the seed set. Therefore, the optimized HyperIM is able to provide ($1-\frac{1}{e}-\epsilon$)-approximate guarantee while reducing the costs of generating the RR sets.

This paper makes the following contributions.
\begin{itemize}
\item  Our in-depth analysis reveals that the state-of-the-art IM algorithms in a hypergraph have low accuracy but large time complexity. To settle this, we propose \emph{HyperIM} that employs stratified sampling to efficiently differentiate the vertices based on their structures in a hypergraph by assigning different sampling probabilities to generate the RR sets.
\item We further develop stratified sampling combined with two sampling
strategies in \emph{HyperIM}, a Binomial-based strategy and a Possion-based strategy,
to increase the sampling efficiency in terms of a smaller number of required samplings and low sampling variances.
\item We provide theoretical analysis to show that \emph{HyperIM} can obtain the seed set with higher influence spread, higher sampling efficiency and lower time complexity. To further cut down the costs of generating RR sets, we more deeply explore the stratified information in \emph{HyperIM} to derive the required number of RR sets for discovering the seed set with the desired theoretical guarantee.
\item Experimental results on real-world hypergraphs show that our proposed algorithms improve the influence spread by up to 2.73X on average over the existing IM algorithms while significantly increasing the sampling efficiency and reducing both the running time and the number of RR sets by orders of magnitude.
\end{itemize} 
\section{Preliminaries}
\section{Preliminaries}
\label{sec:preliminary}
\subsection{Problem definition}\label{problem_definition}
\begin{figure}[hbt!]
\centering
\includegraphics[width=0.31\textwidth]{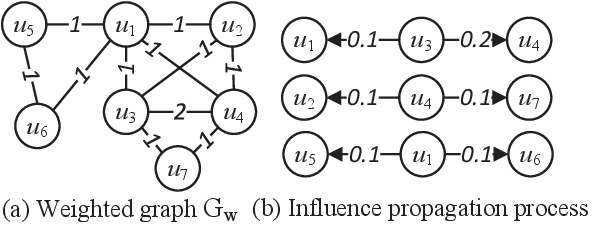}
\caption{Example of an influence propagation process.}\label{propagation-process}
\end{figure}

Let $G=(V,HE)$ denote a hypergraph where $V$ is a vertex set and $HE$ denotes a set of hyperedges.
We use $V(he)$ to denote the set of vertices contained by hyperedge $he$. A hypergraph can be transformed into a regular
graph structurally such that a hyperedge is transformed into several regular edges, where each edge is weighted by the number of hyperedges from which the edge is transformed.
The regular graph transformed from a hypergraph is denoted by $G_w = (V,E,W)$ where $E$ is the edge
set and $W$ is the weight set of the edges.
We use $deg(\mu)$ and $w(\mu,\nu)$ to denote vertex $\mu's$ degree, defined as the number of $\mu's$ neighbors, and the weight of edge $e(\mu,\nu)$ respectively,
where $\mu,\nu\in V$ and $e(\mu,\nu)\in E$.
This paper considers $G_w$ as undirected.
For example, Figure \ref{propagation-process}(a) is the regular graph transformed from the hypergraph $G$ shown in Figure
\ref{fig_example}(b). Table \ref{graphfic_notations} lists notations frequently used in this paper, along with their definitions.
Unless stated otherwise, the words \emph{activate} and \emph{influence} are interchangeably used throughout the paper.
\begin{table}[htb!]
 \centering
  \renewcommand{\baselinestretch}{0.9}\scriptsize
 \caption{Frequently-used notations in this paper}\label{graphfic_notations}
   \begin{tabular}{| @{\hspace{0em}}p{2cm}<{\centering}@{\hspace{0em}} | @{\hspace{0em}}p{6.4cm}<{\centering}@{\hspace{0em}}|}
 \hline
  $G=(V,He)$ & a hypergraph with vertex set $V$ and hyperedge set $He$ \\
 \hline
  $G_{w}=(V,E,W)$ & a weighted graph transformed from $G$ edge set $E$ and weight set $W$ \\
   \hline
   $deg(\mu)$  & the degree of $\mu$ in $G_{w}$ \\
  \hline
   $A(\mu)$  & the vertex set in which vertices can activate $\mu$ \\
   \hline
   $R(\mu)$  & a random reverse researchable set of $\mu$ \\
   \hline
   $\wedge_{R}(S)$  & the coverage of seed set $S$ with respect to $R$ \\
   \hline
   $l_\mu$  & the number of layers of vertices in $A(\mu)$ activating $\mu$ \\
   \hline
   $L_i$  & the vertex set in the $i^{th}$ layer \\
   \hline
   $\wedge_{R}(\mu|S)$  & the coverage of $\mu$ in the seed set $S$ with respect to $R$ \\
  \hline
   $\lambda,\varepsilon$  & any constant $\lambda,\varepsilon>0$ \\
  \hline
   $S$  & a seed set \\
  \hline
  $\mathbb{C}$ & the discrete-time stochastic cascade \\
  \hline
  $\mathbb{I}_{\mathbb{C}}(S)$ & the expected influence of $S$ under $\mathbb{C}$\\
  \hline
  \end{tabular}
\end{table}

Given a hypergraph and its corresponding graph $G_w$ and a seed set $S$ of vertices,
a discrete-time stochastic cascade model works as that:
(i) At timestamp $0$, we set all the vertices in set $S$ as activated and the other vertices inactivated.
(ii) If vertex $\mu$ is activated at timestamp $i$, it has a probability $p(\mu)\in(0,1)$ to activate the vertices in the hyperedges containing $\mu$ at timestamp $i+1$. After time $i+1$, $\mu$ cannot activate any vertex.
(iii) The influence propagation terminates when none of the vertices can be activated in $G_w$.

Let $\mathbb{C}$ denote a set of cascade models and $I_{C}(S)$ denote the number of vertices activated by the seed set $S$ in $G_w$
for a specific influence propagation model $C$ of $\mathbb{C}$. Then, $\mathbb{I}_{\mathbb{C}}(S)=\mathbb{E}_{C\in \mathbb{C}}[I_{C}(S)]$ is also referred to as the expected influence of seed set $S$ under $C$ of $\mathbb{C}$.
Given $S=\{\mu_3\}$, Figure \ref{propagation-process}(b) shows an example of the propagation process over the weighted graph in Figure \ref{propagation-process}(a).
At timestamp 1, $\mu_3$ activates $\mu_1$ and $\mu_4$ with probabilities of $0.1$ and $0.2$ respectively;
At timestamp 2, $\mu_4$ is able to activate $\mu_2$ and $\mu_7$;
At timestamp 3, $\mu_1$ activates $\mu_5$ and $\mu_6$.
Then, the influence propagation process will stop as all the vertices are activated. Thus, we have $I_{C}(S)=7$.

\begin{myDef}\label{aperiodic}
\textbf{Influence maximization in a hypergraph:}
Given an integer $k$ and a cascade model $C$, the influence maximization in hypergraph $G$
is to discover a seed set $S$ with size $k$ which has the largest expected influence
in the transformed graph $G_w$ defined as $S=Max\{\mathbb{I}_{\mathbb{C}}(S_k)\}$, where $S_k$ is a set containing $k$ vertices.
\end{myDef}

\textbf{Cascade models.} Two cascade models are widely used in the existing IM researches: the Independent Cascade (IC) model and the
Linear Threshold (LT) model. These two cascade models, which we adopt to study the influence propagation in hypergraphs,
have different activating processes.
\begin{itemize}
  \item IC model: assume vertex $\mu$ to be activated at timestamp $i$, then $\mu$ has a probability $p(\mu,\nu)<1$ to
  activate its neighbor vertex $\nu$ at timestamp $i+1$.
  \item LT model: for vertex $\mu$, $\sum_{\nu\in Nei(\mu)}p(\mu,\nu)\geq 1$, where $Nei(\mu)$ is the set of $\mu's$ neighbors and
  $p(\nu,\mu)$ is the probability of $\nu$ activating $\mu$. Given a randomly generated parameter $\lambda(\mu)$ from $[0,1]$,
  if $\mu$ is inactivate at timestamp $i$, it will get activated at timestamp $i+1$ only if
  $\sum_{\nu\in ANei(\mu)}p(\mu,\nu)\geq \lambda(\mu)$, where $ANei(\mu)$ is the set of activated neighboring vertices of $\mu$ at timestamp $i$.
\end{itemize}
\subsection{The existing IM algorithms}\label{existing_methods}
Since information spreads over a hypergraph either through hyperedges or vertices,
the existing IM algorithms can be categorized into hyperedge-based algorithms and vertex-based algorithms.

The hyperedge-based IM algorithms consider a hyperedge as a unit of the information propagation such that an activated
vertex is able to influence the vertices in the currently analyzed hyperedge \cite{zhu2018social}.
For example, A MA et al. \cite{ma2022hyper} propose to employ a refinement-based technique to rank the vertices of a hyperedge according to the amount of influence spread requiring iterative computations.
Aktas et al. \cite{aktas2022influence} settle the IM problem from the aspect of discovering similarity among hyperedges by computing the similarity scores between any two hyperedges.
Zhang et al. \cite{zhang2023influence} propose a method to calculate the influence of both vertices and hyperedges
by using message passing equations but without any theoretical explanations on the required guarantee and time complexity.
Due to the various structures of hyperedges, the hyperedge-based IM algorithms usually require significant computations with time complexity expressed as $\sum_{1\le i\le|HE|}\sum_{1\le j\le |HE|}^{i\neq j}|V(he_i)||V(he_j)|$.
Worse still, hyperedge-based IM algorithms do not provide theoretical guarantees about the approximate results and the time complexity because of complicated connections among hyperedges that contain un-uniform numbers of vertices.

The vertex-based IM algorithms consider information spreading from one vertex to another based on their connections, which can provide solutions to the IM problem with theoretical guarantees of the desired approximation and the time complexity. It is for this reason that this paper focuses on the vertex-based IM algorithms. To reduce the time complexity, the state-of-the-art vertex-based IM algorithms mainly employ sampling for generating random reverse researchable (RR) sets to discover the seed set.
A typical sampling method used to produce a random RR set first randomly selects a vertex $\mu$ from $V$ ($\mu\in V$), then
generates the RR set $R(\mu)$ from the set of vertices that can activate $\mu$ with their respective probabilities.
Thus, set $R(\mu)$ is called as a random RR set of $\mu$.
The following equation describes the relationship between the expected influence of seed set $S$ and RR sets labeled as $R$.
\begin{equation}
\renewcommand{\baselinestretch}{1.3}\scriptsize
\begin{aligned}
\label{inf_equation}
&\mathbb{I}_{\mathbb{C}}(S)=|V|\cdot Pr[S\cap R\neq \emptyset]
\end{aligned}
\end{equation}
Equation (\ref{inf_equation}) infers that the probability of the vertices in set $S$ emerging in $R$ is necessary to discover the influence of the seed set $S$.
$\wedge_{R}(S)$ is defined as the coverage of the seed set $S$ that is considered as the number of RR sets covering the seed set $S$.
Thus, $\frac{\wedge_{R}(S)}{|R|}$  can be used to estimate the value of $Pr[S\cap R\neq \emptyset]$ in Equation (\ref{inf_equation})
and the influence of the seed set $S$. For seed set $S$ and vertex $\mu$, we define the marginal coverage of $\mu$ as follows:
\begin{equation}
\renewcommand{\baselinestretch}{1.3}\scriptsize
\begin{aligned}
\label{marginal_cover}
&\wedge_{R}(\mu|S)=\wedge_{R}(\{\mu\cup S\})-\wedge_{R}(S)
\end{aligned}
\end{equation}

Based on Equation (\ref{marginal_cover}), Algorithm \ref{MaxCoverage} is to iteratively choose the vertex with maximum marginal coverage with respect to the current seed set and then the vertex is added into the seed set.
The standard greedy algorithm in Algorithm
\ref{MaxCoverage} lays a foundation for the existing IM algorithms by generating RR sets to discover the seed set.
To reduce the time complexity of the greedy algorithm, the existing IM algorithms focus on the study of the sampling methods
and the number of required RR sets.
\begin{algorithm}
\renewcommand{\baselinestretch}{1}\scriptsize
\caption{Vertex-Selection-Greedy}
\label{MaxCoverage}
$S_{k}\gets \emptyset$\;
Generate $\theta$ random RR sets\;
\For{$i=1 \to i=k$}
{
    identify $v_i$ covering the most RR sets:
    $v_i\gets Max_{\mu\in V}\{\wedge_{R}(\mu|S)\}$\;
    $S_{k}\gets S_{k}\cup v_i$\;
    Remove the RR sets containing $v_i$
}
\end{algorithm}
\subsubsection{Sampling methods for RR set generations}\label{subsubsection21}
Given vertex $\mu$, its random RR set is generated by sampling from set $A(\mu)$ in which vertices can activate $\mu$.
The frequently-used sampling methods for generating random RR sets include reverse influence sampling \cite{tang2015influence, nguyen2016stop, tang2018online} and subset sampling \cite{guo2020influence, yi2023optimal} described as follows.

\textbf{Reverse influence sampling.} Many IM algorithms use reverse influence sampling to generate RR sets \cite{borgs2014maximizing, tang2014influence, tang2015influence}. For each vertex in the sample set $(A(\mu))$ of a given vertex $(\mu)$, reverse influence sampling determines whether it is added into a random RR set by flipping a coin according to its probability of activating $\mu$. The number of the sampling operations is equal to that of vertices in $A(\mu)$. In general, the existing IM algorithms consider the influence probability of vertices in the sample set as the same, i.e., $p(\mu)$. Therefore, the probability distribution of the reverse influence sampling follows the $0-1$ distribution. Thus, the expected number of executing the sampling operations is $|A(\mu)|$ while the expected number of vertices in the RR set is $|A(\mu)|p(\mu)$ with the sampling variance being equal to $|A(\mu)|p(\mu)(1-p(\mu))$.

\textbf{Subset sampling.} Since the reverse influence sampling requires to execute an operation on each vertex in the sample set to determine its addition into a random RR set, the sampling cost is relatively high considering the total sampling operations over a large graph. Wei.et al. \cite{guo2020influence} proposed a subset sampling method to reduce the number of sampling operations. Given vertex $\mu$ and $R(\mu)=\emptyset$, it uses geometric distribution based sampling to iteratively determine the selected position (i.e., $i$) of an element in $A(\mu)$ and then adds the vertex in the $i^{th}$ position of $A(\mu)$ into $R(\mu)$. The probability of the $i^{th}$ vertex in $A(\mu)$ being added into $R(\mu)$
 is expressed as $P(i)=(1-p(\mu))^{i-1}p(\mu)$. Thus, given a random value (labeled as $U$) from $(0,1)$,
 the value of $i$ is computed as: $i=\lceil\frac{\log U}{\log (1-p(\mu))} \rceil$. If $i> |A(\mu)|$,
 it indicates that none of the elements is chosen and the sampling process is terminated. Based on subset sampling,
 Wei et al. \cite{guo2020influence} proposed an IM algorithm called \emph{SUMSIM}. As shown in Line $7$ and Line $14$ of Algorithm \ref{SUBSIM}, \emph{ SUBSIM} is able to skip the un-sampled vertices while directly positioning the sampled vertices.
\begin{algorithm}
\renewcommand{\baselinestretch}{1}\scriptsize
\caption{SUBSIM}
\label{SUBSIM}
Randomly sample a node $\mu\in V$ and set $R_{\mu}=\{\mu\}$\;
Initialize $Q=\{\mu\}$ and $activated(\mu)=true$\;
\While{$Q\neq \emptyset$}
{
$\mu\gets pop(Q)$\;
Let $\mu[t] (t=1,2,...,n)$ be the vertex set activating $\mu$\;
$p(\mu)$ is the probability activating $\mu$\;
$i\gets \lceil\frac{\log U}{\log (1-p)} \rceil$\;
\While {$pos_i\leq n-i$}
{
 $\omega\gets \mu[i]$\;
 \If {$\omega$ is not activated}
 {
    Add $\omega$ to $R$\;
    Add $\omega$ to queue $Q$ \;
    $activated[\omega]\gets$ true\;
 }
 $i\gets i+\gets \lceil\frac{\log U}{\log (1-p)} \rceil$\;
}
Return $R$\;
}
\end{algorithm}

According to Lemma 3 in \cite{guo2020influence}, the expected number of  subset sampling operations required for generating $\mu's$ RR set is $1+|A(\mu)|p(\mu)$. Thus, compared to the reverse influence sampling, the number of sampling operations using \emph{SUBSIM} is greatly reduced. However, \emph{SUBSIM} significantly increases the sampling variance. Based on the knowledge of the geometric sampling, the sampling variance of each geometric sampling is $\frac{1-p(\mu)}{p(\mu)^{2}}$. Since the sampling operations are executed independently, the total sampling variance of generating $R(\mu)$ using Algorithm \ref{SUBSIM} is $\frac{|A(\mu)|p(\mu)(1-p(\mu))}{p(\mu)^{2}}$, which is obviously larger than $|A(\mu)|p(\mu)(1-p(\mu))$ that is the sampling variance of the reverse influence sampling as described above. Even though the vertices in $A(\mu)$ have different probabilities to be added into the RR set, Lemma 5 in \cite{guo2020influence} shows that the probability distribution of the sampling operations still follows the geometric distribution and thus \emph{SUBSIM} also has large sampling variances.

\subsubsection{The number of RR sets}
Borgs et al. \cite{borgs2014maximizing} have proved that the greedy algorithm shown in Algorithm 1 is able to provide a $(1-\frac{1}{e}-\varepsilon)$-approximate solution in $O(k(|V|+|E|)\varepsilon^{-3}\log^{2}|V|)$ time. Tang et al. \cite{tang2014influence, tang2015influence} employ Chernoff bounds \cite{tang2014influence} to further reduce the time complexity by reducing the number of RR sets.

Specifically, Tang et al. \cite{tang2014influence, tang2015influence} propose \emph{Tim+} and \emph{IMM} by employing the Chernoff bounds to derive the number of required RR sets. They have proved that the greedy algorithm is able to provide
an approximate guarantee in $O(k(|V|+|E|)\varepsilon^{-2}\log |V|)$ time. Nguyen et al. \cite{ nguyen2016stop} observe that the time
complexity of \emph{IMM} is highly related to the value of $k$ and it requires a large number of RR sets to be generated with a large $k$. Thus, they propose the Stop-and-Stare algorithm (\emph{SSA}) and the Dynamic Stop-and-Stare algorithm (\emph{D-SSA}) to address the problem. However, as shown in \cite{huang2017revisiting, tang2018online}, \emph{SSA} and \emph{D-SSA} do not provide an approximation guarantee on the influence. Tang et al. \cite{tang2018online} further propose an algorithm, named \emph{OPIM-C} to divide the RR sets into two equal groups for deriving an upper bound and a lower bound by employing the Chernoff bounds to obtain the approximation guarantee of the seed set. Then, the IM algorithm stops as soon as the ratio defined as the lower bound divided by the upper bound is smaller than the desired approximation guarantee. With the bounds, \emph{OPIM-C} is able to provide the desired approximation guarantee while reducing the number of RR sets. The latest IM algorithm based on the generation of RR sets is \emph{SUBSIM}, proposed by Guo et al.\cite{ guo2020influence}. \emph{SUBSIM} reduces the average size of random RR sets by two phases of executions: a sentinel set selection phase and an IM-Sentinel phase. The first phase looks for a seed set with $b, (b<k)$ while the second phase discovers a \emph{(k-b)}-size seed set. The IM-Sentinel phase will stop when the RR set reaches a vertex that has been added into the seed set in the sentinel set selection phase. \emph{SUBSIM} derives the upper bound and the lower bound in the same way as \emph{OPIM-C} to derive the stop condition for the sentinel set selection. Compared to \emph{OPIM-C}, \emph{SUBSIM} does not reduce the required number of RR sets but it reduces the size of each RR set. From these descriptions, we can see that these existing IM algorithms mainly explores the already-generated RR sets to derive the bounds of the influence of the seed set and then obtain the desired approximation based on the bounds.

Based on the above analysis of the existing IM algorithms, they share the following drawbacks when used to solve the IM problem
in a hypergraph. First, the existing IM algorithms consider neighbors of a vertex as a single structure of a graph to form the
sample set to generate a random RR set of the vertex. As a result, they ignore the important structural information associated
with hyperedges in a hypergraph which plays an important role in broadcasting information. Second, they tend to use a sampling
strategy with high sampling variance, resulting in the inaccuracy of the obtained seed set. Third, when generating the RR sets,
they tend to infer the lower and upper bounds for deriving an approximation guarantee by dividing the already-generated RR sets into
two groups without any reasonable explanations. Thus, there is a risk of failing to obtain the required approximation with a small number of RR sets and thus the existing IM algorithms require more RR sets to reach the desired approximation which in turn increase overheads. Therefore, we propose \emph{HyperIM} to efficiently and accurately discover the seed set in hypergraphs as described in the following sections.
\section{HyperIM}\label{hyperim}
This section presents \emph{HyperIM} that designs the activation probability based on the division of the sample set and combines stratified sampling with a Binomial-based strategy and a Possion-based strategy to generate random RR sets in Section \ref{hyperimSampleset}.
Furthermore, we present a theoretical analysis of \emph{HyperIM} in Section \ref{theoretical_bounds}
in terms of influence spread, the sampling efficiency and accuracy and time complexity
while explaining the advantages of \emph{HyperIM}.
\subsection{Stratified sampling for generating RR sets}\label{hyperimSampleset}
\subsubsection{Sample set division.} 
\begin{figure}[hbt!]
\centering
\includegraphics[width=0.30\textwidth]{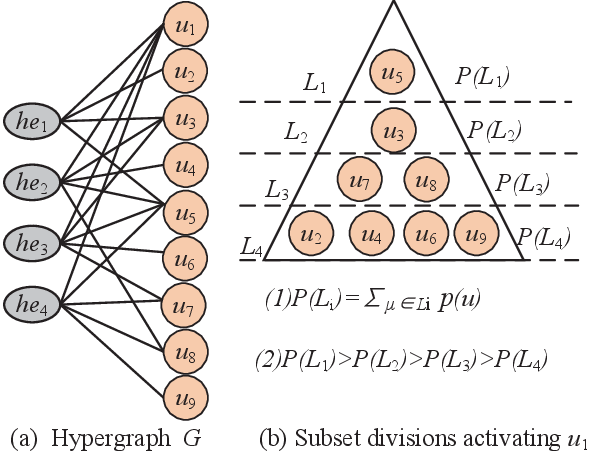}
\caption{An example of sample set divisions according to the vertices' ability to activate $u_1$ in hypergraph $G$.}\label{fig_setdivid}
\end{figure}

Given vertex $\mu$ in hypergraph $G$,
its sample set $A(\mu)$ is formed by the vertices that can activate $\mu$.
The vertices in $A(\mu)$ can be called $\mu's$ neighbors in the corresponding weighted graph $G_w$ transformed from $G$,
that share the same hyperedges with $\mu$.
The more hyperedges that a vertex is contained in, the more channels the vertex has to activate other vertices.
Based on this observation, we divide vertices of $A(\mu)$ into different layers based on the number of channels
through which a vertex can activate other vertices, as measured by the weights on its associated edges.
That is, given $\nu, \alpha\in A(\mu)$, (i) If $w(\mu,\nu)>w(\mu,\alpha)$, then $\nu$ is placed in a higher layer than $\alpha$.
(ii) If $w(\mu,\nu)=w(\mu,\alpha)$, $\nu$ and $\alpha$ are placed in the same layer.
(iii) The vertices in the same layer have the same activation probability while the sum of probabilities of the vertices in a high layer is larger than that in a low layer. For example, a hypergraph is formed by $9$ vertices and $4$ hyperedges as shown in Figure \ref{fig_setdivid}(a). Then, the vertices that can activate $\mu_1$ form a hierarchical group of $4$ layers as shown in Figure \ref{fig_setdivid}(b) such that vertex $\mu_5$ participates in the largest number of hyperedges and it resides in the top layer
while $\mu_2$, $\mu_4$, $\mu_6$ and $\mu_9$ are placed in the bottom layer because they only participate in one hyperedge each.

\textbf{Activation probability.} Let $l_\mu$, $L_i$ and $n(L_{i})$ denote the total number of layers in $A(\mu)$, the vertex set and the number of vertices in the $i^{th}$ layer. Since the activation probability of a vertex in $A(\mu)$ is highly related to its residence layer in the layered sample set, we construct a function to express the connection between the $i^{th}$ layer and the probability of the vertices in the $i^{th}$ layer activating $\mu$ as $P(L_{i})=\frac{1}{i\times \ln (l_\mu+1)}$.
Each vertex in $L_{i}$ has the same chance to activate $\mu$ with the probability
$\frac{P(L_{i})}{n(L_{i})}$ and thus $P(L_{i})=\sum_{\nu\in L_{i}}p(\nu)$ is easily established where $p(\nu)$ denotes $\nu's$ activation probability in $L_{i}$ as shown in Figure \ref{fig_setdivid}(b).
\begin{myLem}\label{myLem-pro}
Given vertex $\mu$ and its two layers denoted as $L_{i}$ and $L_{j}$ in $A(\mu)$ where $0<i<j\leq l_{\mu}$,
we have $P_\mu(L_{i})>P_\mu(L_{j})$ and $\sum_{i=1}^{i=l_\mu}P_\mu(L_{i})\leq 1$.
\end{myLem}
Proof. As $P(L_{i})=\frac{1}{i\times \ln (l_{\mu}+1)}$ and $P(L_{j})=\frac{1}{j\times \ln (l_{\mu}+1)}$ where $i<j$,
it is easy to infer $P(L_{i})>P(L_{j})$ and $\sum_{i=1}^{i=l_{\mu}}P(L_{i})=\frac{1}{\ln (l_{\mu}+1)}(1+\frac{1}{2}+...+\frac{1}{l_{\mu}})\leq 1$. Lemma \ref{myLem-pro} explains that the activation probability is set reasonably and each vertex in the sample set can be properly added into a RR set based on their corresponding layers.
\subsubsection{Sampling methods}
The probability of a vertex being added into a random RR set is equal to that of the vertex activating $\mu$.
Because the probabilities of vertices in $A(\mu)$ activating $\mu$ are different,
the sampling methods designed with equal probability \cite{guo2020influence, hu2023triangular} are ineffective.
A straightforward way is to use the random sampling to generate a random RR set as follows.

\textbf{Random sampling.} The vertices in $A(\mu)$ are sampled one by one for generating $R(\mu)$ when using random sampling. The time complexity of generating a RR set of vertex $\mu$ is $O(deg(\mu))$. Consequently, the time complexity of generating RR sets using the greedy algorithm (shown in Algorithm \ref{MaxCoverage}) is closely related to the number of samplings for generating RR sets which is $\sum_{\mu\in V} O(deg(\mu))$. To reducing the number of samplings, we design a novel stratified sampling with two sampling strategies, reducing the time complexity to $\sum_{\mu\in V} O(l_\mu)$.

\textbf{Stratified sampling.} Given vertex $\mu$, stratified sampling generates a random RR set of vertex $\mu$
by discriminating the vertices of $A(\mu)$ in different layers.
That is, (i) for each layer, i.e., $L_i\ (i>0)$,
vertices are randomly selected from it to form a random subset, labeled as $S_i$;
(ii) the union of the random subsets sampled from all the layers is considered as a random RR set of $\mu$.
The following two lemmas ensure the correctness and validity of employing stratified sampling to generate a random RR set for a given vertex.
\begin{myLem}\label{myLem-numequal}
The total number of possible RR sets generated by stratified sampling is equal to that generated by random sampling.
\end{myLem}
Proof. Suppose the total number of vertices in $A(\mu)$ is $d$ and
these vertices can be divided into $l$ layers, where $n(L_1)+...+n(L_l)=d$ and $n(L_i)$ is the number of vertices in $L_i$.
The size of the vertices in $\mu's$ $RR$ sets ranges from $0$ to $d$. Given any number, i.e., $j,\ 0\leq j\leq d$,
let $n({R_j}^{r})$ and $n({R_j}^{s})$ denote the numbers of \emph{j-size} RR sets generated by random sampling and stratified sampling respectively.
By random sampling, we have $n({R_j}^{r})=C_{d}^{j}$.
By stratified sampling, we have $n({R_j}^{s})=\sum C_{n(L_{1})}^{a_1}\cdot\cdot\cdot C_{n(L_{l})}^{a_l}$,
where $a_1+...+a_l=j$. Given $a$, $b$ and $1\leq a,b\leq l$, according to the Vandermonde's identity which is expressed as $C_{m}^{x}=\sum_{0\leq i\leq m}C_{n(L_{a})}^{i}C_{n(L_{b})}^{x-i}$ \cite{gould1972new} where $m=n(L_{a})+n(L_{b})$ and $0\leq x\leq Min\{n(L_{a}),n(L_{b})\}$, we can easily infer $n({R_j}^{r})=n({R_j}^{s})$ and thus $\sum\limits_{1\leq j\leq d}n({R_j}^{r})=\sum\limits_{1\leq j\leq d}n({R_j}^{s})$.
\begin{myLem}\label{myLem-pronumeq}
The probability of a random RR set being generated by stratified sampling is equal to that by random sampling.
\end{myLem}
Proof. Given two sets $S=\{\mu_1,...,\mu_d\}$, $R=\{\mu_1,...,\mu_s\}$ and $R\subset S$,
let $p(\mu_i) (1\le i\le d)$ denote the probability of vertex $\mu_i$ emerging in $R$.
The probability of generating $R$ from $S$ with random sampling is expressed as $p_{r}(R)=p(\mu_1)\cdot\cdot\cdot p(\mu_s)\cdot (1-p(\mu_{s+1}))\cdot\cdot\cdot(1-p(\mu_d))$.
Suppose $R_i$ denote the subset of $R$ in the $i^{th} (1\leq i\leq l)$ layer,
then $p(R_i)=\prod\limits_{\mu_s\in R_i}p(\mu_s)\cdot\prod\limits_{\mu_t\in (L_i-R_i)}(1-p(\mu_t))$ and the probability of producing $R$ by stratified sampling is
$p_{s}(R)=\prod\limits_{1\leq i\leq r}p(R_i)$. Since $p_{s}(R)$ doesnot change the value of each multiplier in $p_{r}(R)$ as well as the number of multipliers, we have $p_{s}(R)=p_{r}(R)$.
\subsubsection{Sampling strategy}\label{selection-strategy}
A straightforward way to use stratified sampling is to employ the latest sampling strategies, such as the subset sampling \cite{guo2020influence} in each layer to generate RR sets. However, as described in Section \ref{subsubsection21}, the subset sampling has low sampling efficiency and large sampling variance which are fundamentally decided by the distributions of sampling operations.
Therefore, we redesign the sampling strategies to change the distributions of the sampling operations for increasing the sampling efficiency and decreasing the sampling variance. Specifically, when generating a random subset from $L_i$, we tailor design a Binomial-based strategy and a Possion-based strategy by setting a selection condition as that: if $|L_i|\leq 20$, the Binomial-based strategy is used; Otherwise, the Possion-based strategy is used. The distributions of the sampling operations follow the binomial distribution if $|L_i|\leq 20$ or the Poisson distribution if $|L_i|>20$. We describe the two sampling strategies in detail as follows.

$\bullet\ $\textbf{Binomial-based strategy}. If $|L_i|\leq 20$, a binomial-based strategy is used to obtain the specific subset. As elements in the same layer are sampled with equal probability labeled as $p_t$, the probability distribution of selecting a subset whose size ranges from $0$ to $|L_i|$ follows the binomial distribution. Thus, we use $p_t$ and $|L_i|$ to construct a Binomial-based function and for any given number, the function is able to determine its corresponding probability under the Binomial-based distribution and then output a number, i.e., $h$, between $0$ to $|L_i|$ with the probability equaling to $p(h)=C_{|L_i|}^{h}p_t^{h}(1-p_t)^{(|L_i|-h)}$. Therefore, given a number $U$ generated uniformly at random, we use the Binomial-based function to determine the size of a sample subset.

$\bullet\ $\textbf{Possion-based strategy}. When using the Binomial-based strategy, it is required to compute the permutation and combination of the vertices in $L_i$ for determining the probability of a random subset with some size. However, if $|L_i|>20$, such computations are time-cost. The knowledge of the connection between the Binomial distribution and the Possion distribution \cite{hong2013computing, gomez2010another} shows that: when the number of elements in the sample set is larger than $20$ and each element is sampled with a probability that is smaller than $0.05$, the Poisson formula can be leveraged to approximately compute the sampling probability to obtain the size of a random subset. Let $p_f$ denote the sampling probability of an element in $L_i$ and we have $p_f<\frac{1}{20\ln(20+1)}<<0.05$. Thus, the sampling probability of a $h-size$ subset from $L_i$ using the Poisson formula is $p(h)=\frac{\lambda^{h}\cdot e^{-\lambda}}{h!}$ where $\lambda=|L_i|\cdot p_f$. Therefore, when $|L_i|>20$, we propose a Possion-based strategy by constructing a Possion-based function that works as that: for any given number, the function is able to determine its corresponding probability under the Possion-based distribution and then output a number, i.e., $h$, between $0$ to $|L_i|$ with the probability equaling to $p(h)=\frac{\lambda^{h}\cdot e^{-\lambda}}{h!}$.

Both the Binomial-based strategy and the Possion-based strategy are able to determine the size of a RR set, i.e, $h$. Once $h$ is determined, we select $h$ vertices from $L_i$. This process is equivalent to iteratively select one element randomly from the remainder subset which have removed the already-selected elements from $L_i$. Thus, we generate a random integer $a$ and $1\leq a\leq |L_i|-r$ ($0\leq r\leq h-1$) to determine the position of the selected element in the $r^{th}$ iteration. These selected vertices form the \emph{h-size} subset that is consider as a random RR set generated from the sample set $L_i$. The stratified sampling combined with the two sampling strategies leads to the generation algorithms of RR sets as show in Algorithm \ref{HyperIMS}. Firstly, a random number is generated in Line 7. Then, either a Binomial-based function in Line 9 or a Possion-based function in Line 11 is used to determine the size of a RR set and then Lines 13-17 are used to add the vertices into the RR set.
\begin{algorithm}
\renewcommand{\baselinestretch}{1}\scriptsize
\caption{HyperIM}
\label{HyperIMS}
Randomly sample a node $\mu\in V$ and set $R_{\mu}=\{\mu\}$\;
Initialize $Q=\{\mu\}$ and $activated(\mu)=true$\;
\While{$Q\neq \emptyset$}
{
$\mu=Pop(Q)$\;
$L(\mu)=layer(\mu)$\;
\For{$i\leq L(\mu)$}
{
  $Random\_engine(generator)$\;
 \If{$n(L_i)\leq 20$}
 {
    $Dis\gets Binomial\_distribution(n(L_i),p(L_i))$\;
 }
 \Else
 {
    $Dis\gets Poisson\_distribution(1)$\;
 }
  $h \gets Dis(generator)$\;
  \For {$r\leq h$}
  {
     Generate a uniform number randomly from $|L_i|-r$\;
    $R(\mu)\gets$ the selected vertices in $L_i-R(\mu)$\;
     Mark the vertices in $R(\mu)$ as activated\;
    $Q\gets R(\mu)$ and $R\gets R\cup R(\mu)$\;
  }
}
}
\end{algorithm}
\subsection{Theoretical analysis for HyperIM}\label{theoretical_bounds}
We analyze HyperIM from three aspects: (i) influence spread that ultimately decides the number of influential vertices for a \emph{k-size} seed set; (ii) sampling efficiency and accuracy that reflect the quality of the RR sets by sampling used by the IM algorithms; (iii) the time complexity required to reach the desired approximation guarantee.
\subsubsection{Influence spread}
\begin{figure}[hbt!]
\centering
\subfigure[Activation process]{\label{influ_process}
\includegraphics[width=0.21\textwidth]{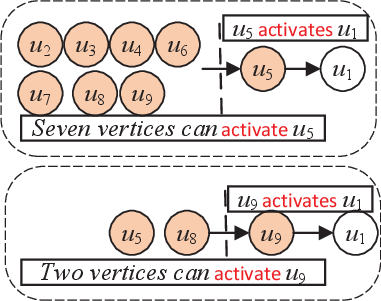}
}\subfigure[Activation probability]{\label{pro_setting}
\includegraphics[width=0.22\textwidth]{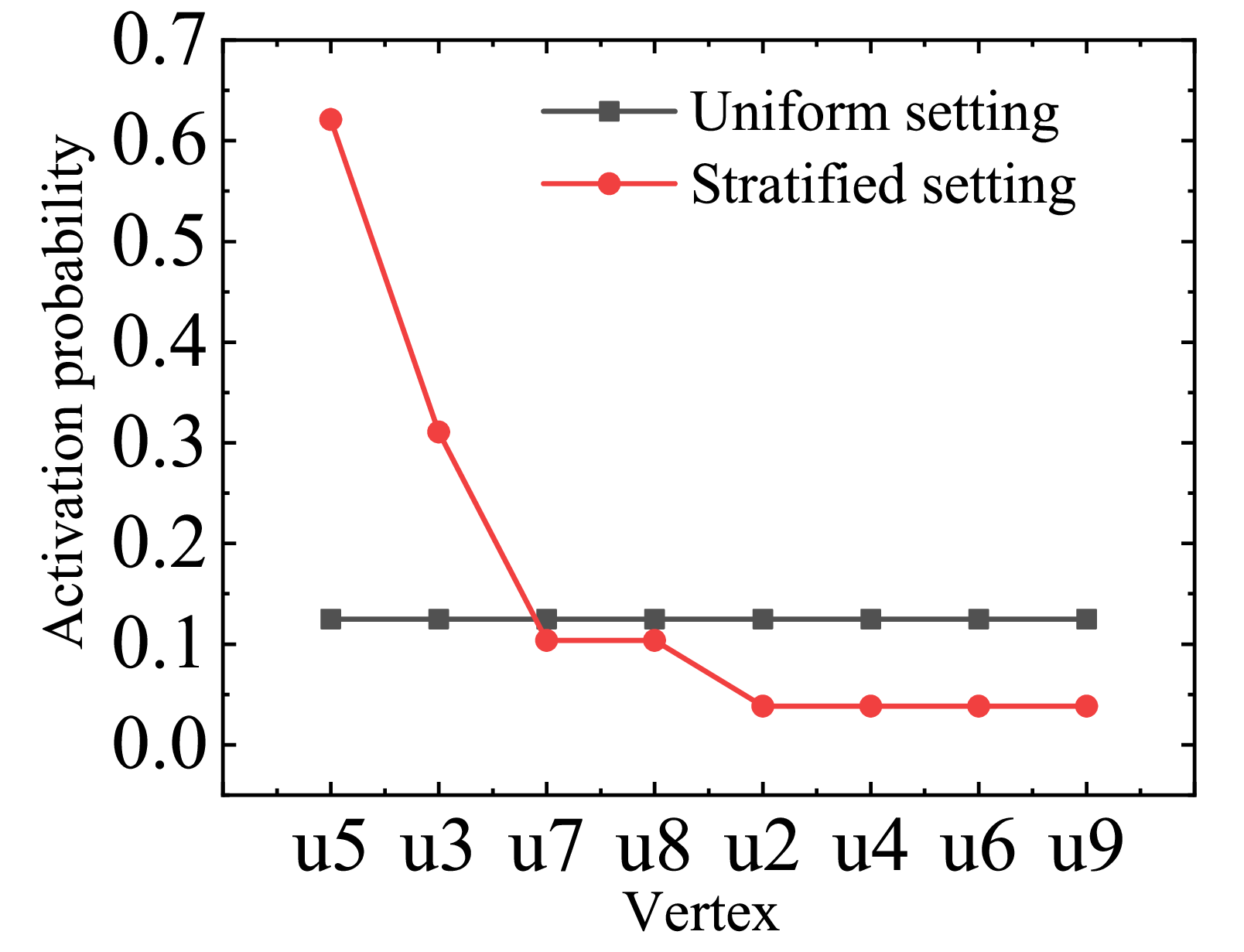}
}\caption{Activation probabilities of vertices in Figure \ref{fig_setdivid}(b) under stratified setting and uniform setting while the stratified setting benefits to obtain influential RR sets.}\label{fig_pro_setting}
\end{figure}

The main step of maximizing the number of influential vertices is to generate the influential RR sets by sampling, meaning that the high influential vertices should be endowed with large probabilities to be added into the RR sets. However, almost all the existing IM algorithms employ the uniform setting, which considers the activation probability of vertices as same in the sample set. Thus, the uniform setting is conducive to obtaining influential RR sets. We set the activation probability differently in \emph{HyperIM} which is called as stratified sampling. As show in Figure \ref{fig_pro_setting}(b), the probabilities of vertices in Figure \ref{fig_pro_setting}(a) to activate $\mu_1$ are different in stratified sampling while they are equal in uniform settings. The stratified setting of the activation probability benefits to obtain the influential RR sets.
For example, since the cascade models (described in Section \ref{problem_definition}) imply that only activated vertices have the ability to influence other vertices, the probabilities of $\mu_5$ and $\mu_9$ activating $\mu_1$ in Figure \ref{fig_setdivid}(a) are positively correlated to the number of vertices activating $\mu_5$ and $\mu_9$ respectively. There are $7$ vertices activating $\mu_5$ but only two vertices can activate $\mu_9$ as shown in Figure \ref{fig_pro_setting}(b). Consequently, the total number of $\mu_5's$ related RR sets is larger than that of $\mu_9's$ relevant RR sets. According to the greedy algorithm shown in Algorithm \ref{MaxCoverage}, $\mu_5$ has larger influence spread than $\mu_9$ and $\mu_5$ is endowed with higher activation probability than $\mu_9$ as shown in \ref{fig_pro_setting}(a). On the other hand, the different numbers of vertices activating $\mu_5$ and $\mu_9$ fundamentally lie in the different number of hyperedges containing $\mu_5$ and $\mu_9$ as show in Figure \ref{fig_setdivid}(a). Therefore, \emph{HyperIM} employs the way of stratified setting to render that the activation probability of a vertex is positively related to the number of its corresponding hyperedges for generating influential RR sets.

We demonstrate the vertices chosen by \emph{HyperIM} with larger influence spread in a formal way. Given vertex $\mu$ and its $t$ hyperedges, i.e., $he_1, he_2,\ ... \ he_t$, the probability of $\mu$ (labeled as $p(\mu)$) emerging in RR sets is related to the number of $\mu's$ hyperedges. The marginal coverage of $\mu$ with respect to seed set $S$ can be expressed as the number of subsets containing $\mu$ as shown in Equation (\ref{marginal_vertex}).
\begin{equation}
\begin{aligned}
\label{marginal_vertex}
&\wedge_{R}(\mu|S)=\wedge_{R}(\{\mu\cup S\})p(\mu)-\wedge_{R}(S)\\
&=\sum_{R(\mu)\subset\{V(he_1)\cup V(he_2)...\cup V(he_t)\}}p(\mu)
\end{aligned}
\end{equation}
Since \emph{HyperIM} tends to select the vertices with high activation probability and these vertices are usually contained in a large number of hyperedges, the marginal coverage of the selected vertices computed by Equation (\ref{marginal_vertex}) is certainly large. Therefore, the seed set that is selected from the generated RR sets has higher influence spread than that produced by the existing IM methods which treat the vertices equally when to select the vertices for generating the RR sets.
\subsubsection{Sampling efficiency and accuracy}
When to use sampling methods to generate RR sets, the sampling efficiency and the sampling variance are important metrics to evaluate the IM algorithms. The sampling efficiency is measured by the number of samplings required to obtain the expected size of the RR sets. The sampling variance is to evaluate the accuracy of the sampling methods. The larger sampling variance means a larger number of RR sets required to reduce the estimated errors. Before comparing the sampling efficiency and accuracy to the sampling methods used in the existing IM algorithms, we analyze the expected size of a RR set and the required number of sampling operations in \emph{HyperIM} as shown in Theorem \ref{size-sample}.
\begin{myTheo}\label{size-sample}
Given a sample set in which the vertices are divided into $l$ $(l\geq 1)$ layers, the expected size of a random RR set labeled as $R$ is $1+\frac{l}{\ln (l+1)}$ with the number of sampling operations equaling to $l$ while
the sampling variance ranges from $\frac{l-\sum_{i=1}^{i=l}p_i}{\ln(l+1)}$ to $\frac{l}{\ln (l+1)}$, where $p_i$ denotes
the sampling probability of a vertex in the $i^{th}$ layer to be added into $R$.
\end{myTheo}
Proof. Let $E(R)$ and $V(R)$ denote the expected size and the sampling variance of set $R$ generated by sampling.
The sampling operations in different layers are independent so that we have $E(R)=\sum_{i=1}^{i=l}E(R_i)$
and $V(R)=\sum_{i=1}^{i=l}V(R_i)$, where $R_i$ denotes a random RR subset of $L_i$.
When $|L_i|\leq 20$, the probability distribution of obtaining $R_i$ from $L_i$ follows the binomial distribution and thus the expected size of $R_i$ is $|L_i|p_i$ and the
sampling variance is $|L_i|p_i(1-p_i)=\sum_{i=1}^{i=l}(1-p_i)|L_i|\frac{1}{|L_i|\ln(l+1)}=\frac{l-\sum_{i=1}^{i=l}p_i}{\ln(l+1)}$. When $|L_i|>20$, the probability distribution for obtaining $R_i$ follows the Poisson distribution. Thus,
both the expected size and the sampling variance for obtaining $R_i$ are $|L_i|p_i$. Furthermore, we need to add at least one vertex into a RR set and thus we have $E(R)=1+\sum_{i=1}^{i=l}|L_i|p_i=1+\sum_{i=1}^{i=l}|L_i|\frac{1}{|L_i|\ln(l+1)}=1+\frac{l}{\ln (l+1)}$ and $\frac{l-\sum_{i=1}^{i=l}p_i}{\ln(l+1)}\leq V(R)\leq\frac{l}{\ln (l+1)}$.
As described in Section \ref{selection-strategy}, both the Binomial-based strategy and the Possion-based strategy just requires one execution in each layer to determine the size of a RR set. Therefore, the number of sampling operations is equal to the number of layers of the sample sets.

\textbf{\emph{Comparisons to the existing IM algorithms.}}
Since \emph{SUMSIM}\cite{guo2020influence} represents the existing sophisticated IM algorithm, we compare \emph{HyperIM} with \emph{SUMSIM} theoretically from the aspects of both sampling efficiency and sampling variances as follows.

As the activation probabilities in the same layer are equal, \emph{SUMSIM} can easily execute the subset sampling in each layer. Suppose $l$ be the total number of layers and $p_i$ denote the sampling probability of a vertex being added into a random RR set from $L_i$ whose number of vertices is labeled as $n_i$ where $1\leq i\leq l$. According to Lemma $3$ in \cite{guo2020influence}, the expected size of a RR set produced by \emph{SUBSIM} is $1+\sum_{i=1}^{i=l}n_ip_i=1+\frac{l}{\ln(l+1)}$ which requires $|L_i|$ samplings. As the subset sampling follows the geometric distribution, the sampling variance is $\sum_{i=1}^{i=l}\frac{n_ip_i(1-p_i)}{p_i^2}$. Therefore, when required to reach the expected size of a RR set at the $i^{th}$ layer, \emph{HyperIM} requires to execute one sampling while \emph{SUMSIM} requires $|L_i|$ samplings. On the other hand, at the $i^{th}$ layer, the sampling variance in \emph{HyperIM} can be reduced to $n_ip_i(1-p_i)$ in binomial-based strategy and $n_ip_i$ in Possion-based sampling strategy which is much smaller than $\frac{n_ip_i(1-p_i)}{p_i^2}$ caused by \emph{SUMSIM}. Table \ref{comparison-1} shows a brief comparison between \emph{HyperIM} and \emph{SUBSIM}. Table \ref{comparison-1} presents that when to generate the same size of RR sets, \emph{SUMSIM} requires a much larger number of sampling operations than \emph{HyperIM} and meanwhile \emph{SUMSIM} leads to much larger sampling variances. This is because the sampling distribution in \emph{HyperIM} follows the Binomial distribution or the Poisson Distribution that fundamentally decides the high sampling efficiency and low sampling variance.
\begin{table}[htb!]
\centering
  \renewcommand{\baselinestretch}{1.3}\scriptsize
 \caption{The comparisons between \emph{HyperIM} and \emph{SUBSIM}.}\label{comparison-1}
   \begin{tabular}{|@{\hspace{0em}}p{1.2cm}<{\centering}@{\hspace{0em}}|@{\hspace{0em}}p{2cm}<{\centering}@{\hspace{0em}}|@{\hspace{0em}}p{1.35cm}<{\centering}@{\hspace{0em}}|@{\hspace{0em}}p{1.75cm}<{\centering}@{\hspace{0em}}|@{\hspace{0em}}p{2.2cm}<{\centering}@{\hspace{0em}}|}
  \hline
   IM algorithm & Probability distribution & Expected size&Expected samplings & Sampling variance  \\
  \hline
   SUBSIM \cite{guo2020influence}  &Geometric distribution &$1+\frac{l}{\ln(l+1)}$ &$\sum_{i=1}^{i=l}|L_i|$ &$\sum_{i=1}^{i=l}\frac{n_ip_i(1-p_i)}{p_i^2}$  \\
  \hline
  \multirow{2}{*}{HyperIM} &Binomial distribution &$1+\frac{l}{\ln(l+1)}$  & $l$ & $\frac{l-\sum_{i=1}^{i=l}p_i}{\ln(l+1)}$ \\

                           &Poisson Distribution &$1+\frac{l}{ln(l+1)}$  & $l$ & $\frac{l}{\ln(l+1)}$ \\

   \hline
  \end{tabular}
\end{table}
\subsubsection{Time complexity} This subsection first analyzes the time cost of generating a random RR set and then analyze the total time complexity of \emph{HyperIM}.
\begin{myTheo}\label{mythe-generate-a-RR-set-cost}
Under the IC model and the LT model, for any vertex $\mu$,
the time cost of generating a random RR set can be bounded
by $O(\theta(l^{*})\mathbb{I}_{\mathbb{C}}(\{\mu\})$), where $\theta(l^{*})=\frac{1}{|V|}\sum_{\nu\in V}\frac{l_{\nu}}{\ln (l_{\nu}+1)}$.
\end{myTheo}
Proof. We prove Theorem \ref{mythe-generate-a-RR-set-cost} in two steps:
(i) We firstly describe the time cost of generating a random RR set that should contain a given vertex, i.e., $\mu$;
(ii) Based on the result of the first step, we infer the expected cost to produce a random RR set in a hypergraph.
The following explanations of the two steps are described under the IC model.
The proof under the LT model is similar to that under the IC model so that we omit it for space limits.

Let $Pr[\mu\xrightarrow{L_i}\nu]$ denote the probability of $\mu$ in the $i^{th}$ layer of the sample set activating $\nu$.
Thus, the probability of $\mu$ emerging in $\nu's$ RR set is $Pr[\mu\xrightarrow{L_i}\nu]$.
According to the cascade model, only if the vertices activated at the previous timestamp have the ability to influence the other vertices,
we should evaluate the expected chances of the vertices activating $\mu$ at the current timestamp to ensure that $\mu$ has a chance to activate $\nu$ at the next timestamp and thus to be added into $\nu's$ RR set.
Let $Pr[(\omega,\mu)\xrightarrow{L_i}\nu]$ be the probability of $\mu$ activating $\nu$ through $\omega$ and $p(\omega,\mu)$ be the probability of
$\omega$ activating $\mu$. Therefore, we have:
\begin{equation}
\renewcommand{\baselinestretch}{1.3}\scriptsize
\begin{aligned}
\label{pro_vinfection}
&Pr[(\omega,\mu)\xrightarrow{L_i}\nu]=p(\omega,\mu)\times Pr[\mu\xrightarrow{L_i}\nu]
\end{aligned}
\end{equation}

Given vertex $\nu$, the expected cost of generating $\nu's$ random RR set containing $\mu$,
denoted as $\mathbb{E}[R_{\mu}^{\nu}]$, is described as below:
\begin{equation}
\renewcommand{\baselinestretch}{1.3}\scriptsize
\begin{aligned}
\label{pro_v_RRset}
&\mathbb{E}[R_{\mu}^{\nu}]=\sum_{\omega\in V, 1\leq i\leq l_{\nu}}Pr[(\omega,\mu)\xrightarrow{L_i}\nu]\\
&=\sum_{\omega\in V,1\leq i\leq l_{\nu}}p(\omega,\mu)\times Pr[\mu\xrightarrow{L_i}\nu]\\
&\leq\sum_{1\leq i\leq l_{\nu}}Pr[\mu\xrightarrow{L_i}\nu]\times deg(\mu)p(\omega,\mu)<<\frac{l_{\nu}\cdot deg(\mu)}{\ln (l_{\nu}+1)}p(\omega,\mu) \\
\end{aligned}
\end{equation}
Let $\mathbb{E}[R]$ denote the expected cost of generating a RR set for any vertex. Since $\theta(l^{*})=\frac{1}{|V|}\sum_{\nu\in V}\frac{l_{\nu}}{\ln (l_{\nu}+1)}$, we have the following equation:
\begin{equation}
\renewcommand{\baselinestretch}{1.3}\scriptsize
\begin{aligned}
\label{pro_whole_RRset}
&\mathbb{E}[R]=\frac{1}{|V|}\sum_{\mu,\nu\in V}\mathbb{E}[R_{\mu}^{\nu}]<<\frac{1}{|V|}\sum_{\mu,\nu\in V}\frac{l_{\nu}\cdot deg(\mu)}{\ln (l_{\nu}+1)}p(\mu,\omega)\\
&=\frac{1}{|V|}\sum_{\nu\in V}\frac{l_{\nu}}{\ln (l_{\nu}+1)}\sum_{\mu\in V}deg(\mu)p(\mu,\omega)\\
&<\frac{1}{|V|}\sum_{\nu\in V}\frac{l_{\nu}}{\ln (l_{\nu}+1)}\mathbb{I}_{\mathbb{C}}(\{\mu\})=\theta(l^{*})\mathbb{I}_{\mathbb{C}}(\{\mu\})
\end{aligned}
\end{equation}
\begin{myTheo}\label{mythe-time-complexity}
Under the IC model and the LT model, the time complexity of \emph{HyperIM} can be bounded by $O(k\cdot \theta(l^{*})\cdot|V|\cdot\log\frac{|V|}{\epsilon^{2}})$ for discovering
a seed set with size $k$ to provide a $(1-\frac{1}{e}-\epsilon)-$approximate solution with $1-\frac{1}{|V|}$ probability.
\end{myTheo}
Proof. Let $OPT_{k}$ define the largest expected influence of the vertex among all the seed sets.
As explained in \cite{tang2015influence, guo2020influence},
$OPT_{k}$ can be used to bound the number of the RR sets as $O(\frac{k\cdot |V|\log |V|}{OPT_{k}\cdot \epsilon^{2}})$ to
provide a $(1-\frac{1}{e}-\epsilon)-$approximate solution with $1-\frac{1}{|V|}$ probability.
As $\mathbb{I}_{\mathbb{C}}(\{\mu\})<OPT_{k}$, we have $\mathbb{E}[R]\leq\theta(l^{*})\cdot OPT_{k}$.
Thus, the time complexity of \emph{HyperIM} can be expressed as follows:
$$O(\frac{k\cdot |V|\log |V|}{OPT_{k}\cdot \epsilon^{2}}\cdot \mathbb{E}[R])=O(k\cdot\theta(l^{*})\cdot |V| \cdot\frac{\log |V|}{\epsilon^{2}}).$$

\textbf{\emph{Comparisons to the existing IM algorithms.}} Among the recent IM algorithms, \emph{SUBSIM} has the least time complexity.
Let $T_{S}$ and $T_{H}$ denote the time complexities of \emph{SUMSIM} and \emph{HyperIM} respectively.
According to Theorem 1 in \cite{guo2020influence} and Theorem \ref{mythe-time-complexity} described above,
$T_{S}$ and $T_{H}$ can be expressed as follows:
$$T_{S}=\theta(\frac{\sum_{\nu\in V}\frac{l_{\nu}}{\ln (l_{\nu}+1)}}{|V|})\cdot T,T_{H}=\theta(\frac{\sum_{\nu\in V} deg(\nu)}{|V|})\cdot T,$$
where $\theta$ is a monotonically increasing concave function and $T=k\cdot |V| \log\frac{|V|}{\epsilon^{2}}$. As $l_{\nu}$ is the number of layers of the sample set whose size is equal to $deg(\nu)$, we have $l_{\nu}<deg(\nu)$ and thus $T_{S}<T_{H}$.
\section{Optimizing HyperIM}
\label{sec:optimization}
With the increase of a hypergraph, the number of RR sets gets large and thus the generation of the RR sets will
cause huge costs as shown in Theorem \ref{mythe-generate-a-RR-set-cost}. This section optimizes \emph{HyperIM} from the aspect of reducing the number of RR sets. We first divide the RR sets by fully employing the stratified information of the sample sets. Then, according to the division, we derive the upper bound and the lower bound to infer the required number of RR sets to obtain the desired approximation of a seed set.
\subsection{Division of the RR sets}
Based on the sample set division and the activation probability in \emph{HyperIM} described in Section \ref{hyperim}, the vertices in high layers usually have larger influential ability than those vertices in low layers. Thus, when generating the RR sets, we propose to contain more vertices of the low layers to derive a tighten upper bound of the influence spread while increasing the number of vertices in high layers to raise the lower bound, leading to the acceleration of shortening the distance between the upper bound and the lower bound. Therefore, we speed the greedy algorithm to discover a seed set by quickly approaching the desired approximation.

Specifically, the random RR sets can be divided into two groups: $R_1$ and $R_2$.
Suppose $F_{cnt}$, $T_{cnt}$ and $L_{cnt}$ be the number of the vertices in the first layers, the total size of the vertices in all the layers and the number of layers of the sample sets. Thus, we use $\alpha=\frac{F_{cnt}}{T_{cnt}}$ to denote the ratio of the vertices in the first layers among those in all the layers and employ $\beta=\frac{L_{cnt}}{|V|}$ to express the average number of layers of a sample set. Then, we use $\alpha$ and $\beta$ to adjust the RR sets in $R_1$ and $R_2$. Let $i$ denote the order of the layer in which the vertices are to be dealt with where $1\leq i\leq \beta$. When $|L_i|<\alpha|A(\mu)|$, $\mu's$ RR set in $R_1$ must contain the vertices in $|L_i|$. Otherwise, we insert $\mu's$ RR set into $R_2$. Then, we use $R_1$ and $R_2$ to derive the upper and the lower bounds and the required number of the RR sets for reaching the desired approximation described by the following subsections.
\subsection{Derivations of the upper and lower bounds}
With respect to seed set $S$,
let $\wedge_{R_1}(\mu|S)$ be the marginal coverage of vertex $\mu$ in $R_1$ and $maxC(S,k)$
represent the set of $k$ vertices with the largest coverage in $R_1$.
Let $p_{max}$ denote the maximum probability of a vertex emerging in $R_1$ which will be certainly added into the optimal set $S^{o}$ defined as the seed set with the maximum influence spread among all the seed sets with size $k$.
Let $\wedge_{R_1}(S^{o})$ denote the marginal coverage of the optimal seed set $S^{o}$. Thus, we bound the upper bound of $\wedge_{R_1}(S^{o})$ as follows.
\begin{myLem}\label{myLem-up}
Given any seed set $S$, we have
$\wedge_{R_1}(S^{o})\leq\wedge_{R_1}(S)+\sum\limits_{\mu\in maxC(S,k)}\wedge_{R_1}(\mu|S)\cdot p_{max}$, where $\wedge_{R_1}(S)$ denotes
the marginal coverage of set $S$.
\end{myLem}
Proof. If $S=S^{o}$, the lemma is obviously valid. Otherwise, at least one vertex in $S$ is not included in $S^{o}$.
 Since $S^{o}$ is the optimal set with the largest influence
and any vertex in $R_1$ can be included in $S$ with a probability, we have the following description.
\begin{equation}
\renewcommand{\baselinestretch}{1.3}\scriptsize
\begin{aligned}
\label{upper}
&\wedge_{R_1}(S^{o})\leq \wedge_{R_1}(S\cup S^{o})\leq\wedge_{R_1}(S)+\sum\limits_{\mu\in maxC(S,k)}\wedge_{R_1}(\mu|S)\cdot p_{max}
\end{aligned}
\end{equation}

Assume $S_{i}^{*}$ be a set with $i$ vertices obtained by the first $i$ iterations of the greedy algorithm.
Based on Lemma \ref{myLem-up}, we derive a upper bound labeled as $\wedge_{R_1}^{u}(S^{o})$ in the following equation:
\begin{equation}
\begin{aligned}
\label{upper_2}
&\wedge_{R_1}^{u}(S^{o})=\min\limits_{0\leq i\leq k}(\wedge_{R_1}(S_{i}^{*})\\
&+\sum\limits_{\mu\in maxC(S_{i}^{*},k)}\wedge_{R_1}(\mu|S_{i}^{*})\cdot p_{max})
\end{aligned}
\end{equation}
Thus, we deduce $\wedge_{R_1}^{u}(S^{o})$ using the following lemma.
\begin{myLem}\label{myLem-5}
$\wedge_{R_1}^{u}(S^{o})<\frac{\wedge_{R_1}(S_{k}^{*})}{1-(1-\frac{1}{k p_{max}})^{k}}$.
\end{myLem}
Proof. Considering the basic nature of the greedy algorithm, given a constant $0\leq i\leq k-1$, we have:
$$\wedge_{R_1}(S_{i+1}^{*})-\wedge_{R_1}(S_{i}^{*})=\max\limits_{\mu\in V}(p_{max}\wedge_{R_1}(\mu|S_{i}^{*})).$$
Thus,
\begin{equation}
\renewcommand{\baselinestretch}{1.3}\scriptsize
\begin{aligned}
\label{upper_4}
&\wedge_{R_1}(S_{i}^{*})+\sum\limits_{\mu\in maxC(S_{i}^{*},k)}p_{max}\wedge_{R_1}(\mu|S_{i}^{*})\\
&\leq\wedge_{R_1}(S_{i}^{*})+k p_{max}\cdot(\wedge_{R_1}(S_{i+1}^{*})-\wedge_{R_1}(S_{i}^{*}))\\
&\wedge_{R_1}(S^{o})-\wedge_{R_1}(S_{i+1}^{*})\leq (1-\frac{1}{k p_{max}})(\wedge_{R_1}(S^{o})-\wedge_{R_1}(S_{i}^{*}))
\end{aligned}
\end{equation}
Recursively, we obtain the following inequality:
\begin{equation}
\renewcommand{\baselinestretch}{1.3}\scriptsize
\begin{aligned}
\label{upper_5}
&\wedge_{R_1}(S^{o})-\wedge_{R_1}(S_{k}^{*})\leq (1-\frac{1}{k p_{max}})^{k}(\wedge_{R_1}(S^{o})-\wedge_{R_1}(S_{o}^{*}))\\
&\leq(1-\frac{1}{k p_{max}})^{k}\wedge_{R_1}(S^{o}).
\end{aligned}
\end{equation}
Therefore, we have the following expression:
\begin{equation}
\renewcommand{\baselinestretch}{1.3}\scriptsize
\begin{aligned}
\label{upper_6}
&\wedge_{R_1}^{u}(S^{o})\leq\frac{\wedge_{R_1}(S_{k}^{*})}{1-(1-\frac{1}{k p_{max}})^{k}}<\frac{\wedge_{R_1}(S_{k}^{*})}{1-(\frac{1}{e})}.
\end{aligned}
\end{equation}
According to lemma 5.2 in \cite{tang2018online}, given a constant $\delta\in (0,1)$ related to the approximation guarantee, we use $R_2$ to derive the lower bound labeled as $\wedge_{R_2}^{l}(S^{o})$ as follows:
\begin{equation}
\renewcommand{\baselinestretch}{1.3}\scriptsize
\begin{aligned}
\label{upper}
&\wedge_{R_2}^{l}(S^{o})=((\sqrt{\wedge_{R_2}(S^{*}_{k})+\frac{2ln\frac{1}{\delta}}{9}}-\sqrt{\frac{ln\frac{1}{\delta}}{2}})^{2}-\frac{ln\frac{1}{\delta}}{18})\cdot\frac{n}{|R_2|}.
\end{aligned}
\end{equation}
\subsection{Determination of the number of RR sets}
\begin{myLem}\label{myLem-4}
Given $\varepsilon,\delta\in(0,1)$, if $|R_1|$ satisfies
$$|R_1|\geq \frac{2|V|(t\sqrt{\ln\frac{2}{\delta}}+\sqrt{t\ln\binom{|V|}{k}}+\ln\frac{2}{\delta})^{2}}{\varepsilon^{2}\cdot \mathbb{I}_{\mathbb{C}}(S^{o}_{k})},$$
where $t=1-\frac{1}{e}$. Then, with at least $1-\delta$ probability,
$$\mathbb{I}_{\mathbb{C}}(S^{*}_{k})\geq (t-\varepsilon)\mathbb{I}_{\mathbb{C}}(S^{o}_{k}).$$
\end{myLem}
The proof of Lemma \ref{myLem-4} is similar to the proofs of Lemma 6.1 in \cite{tang2018online} and Lemma 6 in \cite{guo2020influence}.
Lemma \ref{myLem-4} is used to deduce an upper bound for $|R_1|$, labeled as $\theta_{max}$ as that:
\begin{equation}\label{equation-max}
\theta_{max}=\frac{2|V|(\sqrt{\ln\frac{2}{\varepsilon}}+\sqrt{\ln\binom{|V|}{k}+\ln\frac{6}{\varepsilon}})}{\varepsilon^{2}\cdot k}
\end{equation}
The number of RR sets in $R_1$ and $R_2$ is initially set as $\theta_0=\frac{\theta_{max}\cdot\varepsilon^{2}\cdot k}{|V|}$.
Once we obtain the RR sets including both $R_1$ and $R_2$,
based on the upper bound and the lower bound shown in Equation (\ref{upper_6})and Equation (\ref{upper}) respectively, we compute $\alpha=\wedge_{R_1}^{l}(S^{o})/\wedge_{R_2}^{u}(S^{o})$
as the approximation guarantee of the obtained seed set.

According to the division of RR sets, the inferred upper and lower bounds of the influence spread and the required number of the RR sets, we proposed an optimized algorithm, called \emph{HyperIM\_BRR}, shown in Algorithm \ref{HyperIM_set_Bounds}.
If $\alpha\leq 1-\frac{1}{e}-\varepsilon$, we terminate the algorithm and
return the seed set. Otherwise, we iteratively double the number of the current size of $R_1$ and $R_2$ and
insert new elements to $R_1$ and $R_2$ to obtain a new seed set until $\alpha\leq 1-\frac{1}{e}-\varepsilon$.
As described in \cite{tang2018online}, when the number of the iterations reaches $\lceil\log_{2}\frac{\theta_{max}}{R_0}\rceil$, the algorithm must return a seed set with the required approximation.
\begin{algorithm}
\renewcommand{\baselinestretch}{1}\scriptsize
\caption{HyperIM\_BRR}
\label{HyperIM_set_Bounds}
Initialize $\theta_{max}$ by Equation (\ref{equation-max})\;
$\theta_{0}=\frac{k\theta_{max}\varepsilon}{|V|}$\;
\While{$i\neq \theta_{0}$}
{
Generate $R_1$ and $R_2$\;
$\wedge_{R_1}^{u}(S^{o})=lowerbound(R_1)$\;
$\wedge_{R_2}^{l}(S^{o})=upperbound(R_2)$\;
$\alpha=\wedge_{R_2}^{l}(S^{o})/\wedge_{R_1}^{u}(S^{o})$\;
\If {$\alpha\leq 1-\varepsilon-\frac{1}{e}$}
{Return S\;}
\Else
{Double $\theta_{0}$\;}

}
\end{algorithm}
\section{Related work}
In this paper, we settle the IM problem in a hypergraph by following the basic idea of vertex-based IM algorithms, that is, to study information diffusion among vertices. A large body of researches belong to the vertex-based IM algorithms. Domingos and Richardson \cite{domingos2001mining} pioneer the research of the IM problem. Then, Kempe et al.\cite{kempe2003maximizing} formalize influence maximization as a NP-hard problem and develop a popular greedy algorithm to settle the problem. Since then, lots of research \cite{budak2011limiting, chen2010scalable, chen2009efficient, cheng2014imrank, cohen2014sketch, goyal2011data, goyal2011celf++, goyal2011simpath, jung2012irie, lei2015online, li2015tan} has been proposed to settle the IM problem. Most of these algorithms develop heuristic strategies but without providing the approximation guarantee for discovering a seed set. Instead, a family of RR sets based research \cite{nguyen2016stop, huang2017revisiting, tang2018online, guo2020influence}, as discussed in Section \ref{existing_methods}, are able to efficiently settle the IM problem with a strong theoretical guarantee. These existing vertex-based IM algorithms settle the IM problem in a regular graph and however the important structural information among hyperedges is ignored. Our IM algorithms consider the structural information of hyperedges while improving the efficiency and accuracy to discover an optimal seed set in a hypergraph.

Besides, lots of research work studies the maximum influence under given conditions. For instance, the budgeted influence maximization takes cost into consideration \cite{guo2023efficient, pham2018maximizing}.
Topic-aware IM explore the properties of the topics for propagation \cite{li2015tan, chen2015online} to discover the seed set. Time-aware IM \cite{gomez2011uncovering, liu2012time} considers that the diffusion process in social network is changing over time while competitive IM \cite{budak2011limiting, lu2015competition} studies the influence of social network in face that several competitors prompt their products simultaneously. These studies can be cosidered orthogonal to our study.

\section{Evaluation}
In this section, we present the evaluation of \emph{HyperIM} by extensive experiments conducted on a machine with Intel(R) Xeon(R) Silver 4110 CPU @2.10GHz and 32GB memory.

\textbf{\emph{Datasets.}} We evaluate the IM algorithms on five real-world hypergraph datasets as shown in Table \ref{graphfic_datasets}.
These hypergraphs model the complicated connection among data in different applications \cite{ha2018efficient, klimm2020hypergraphs}. Hypergraph \emph{email-Eu-full} represents data of a mail system where a vertex denotes a sender or a receiver and a hyperedge represents an email including both senders and receivers. Hypergraph \emph{tags-ask-ubuntu} is to model connection among questions and tags where a vertex is a tag and a hyperedge represents a question containing different tags. \emph{NDC-substances-full} models the temporal higher-order network of substances where a hyperedge is to represent a simplex forming by a set of substances. Hypergraph \emph{threads-ask-ubuntu} models the question system where each hyperedge is a question along with its corresponding answers. Hypergraph \emph{coauth-MAG-Geology-full} is to model information of publications where a hyperedge is to model the authors of a publication.
\begin{table}[htb!]
\centering
  \renewcommand{\baselinestretch}{0.9}\scriptsize
 \caption{Summary of hypergraph datasets, where \emph{AVGDeg} denotes the average number of neighbors of a vertex in the weighted graph
 and \emph{AVGSize} is the average number of hyperedges that a vertex participates in.}\label{graphfic_datasets}
   \begin{tabular}{|@{\hspace{0em}}p{2.6cm}<{\centering}@{\hspace{0em}} |@{\hspace{0em}}p{1.1cm}<{\centering}@{\hspace{0em}}|@{\hspace{0em}}p{1.1cm}<{\centering}@{\hspace{0em}}|@{\hspace{0em}}p{1.4cm}<{\centering}@{\hspace{0em}}|@{\hspace{0em}}p{1.8cm}<{\centering}@{\hspace{0em}}|}
  \hline
   Hypergraph & $\arrowvert V \arrowvert$ & $\arrowvert He \arrowvert$ & $\arrowvert E \arrowvert$ & \emph{AVGDeg}$,\ $ \emph{AVGSize}  \\
  \hline
 email-Eu-full  & 1,005 & 25,148 & 66,672 & 132.68$,\ $25.02 \\
  \hline
 tags-ask-ubuntu  &3,029 &147,222 &265,406 &175.24$,\ $48.6  \\
   \hline
 NDC-substances-full &5,556 & 10,273 & 336,826 & 121.24$,\ $1.8 \\
  \hline
 threads-ask-ubuntu  &125,602 &166,999 &374,314 &5.96$,\ $1.32  \\
 \hline
 coauth-MAG-Geology-full  &1,261,129 &1,204,704 &11,165,572 &17.71$,\ $0.95  \\
  \hline
  \end{tabular}
\end{table}

\textbf{\emph{Evaluation methods.}}
To reduce the costs while providing the guarantee as described in Section \ref{existing_methods}, the existing methods mainly
employ the reverse influence sampling and subset sampling to generate the RR sets for discovering the seed set.
We select the latest algorithms to evaluate our proposed algorithms. The first is  \textbf{\emph{SUBSIM \cite{guo2020influence}}} which employs the subset sampling to reduce the generation costs of RR sets. The second is \textbf{\emph{TriangleIM \cite{hu2023triangular}}} which  mainly produces a subset of $k$ vertices with the maximum triangular structural stability defined in \cite{hu2023triangular} by employing the triangle-based edge sampling. In nature, \textbf{\emph{TriangleIM \cite{hu2023triangular}}} follows the basic idea of reverse influence sampling. In contrast, our proposed algorithms of \textbf{\emph{HyperIM}} and \textbf{\emph{HyperIM\_BRR}} use the stratified sampling combined with a Binomial-based strategy and a Possion-based strategy to generate RR sets. Furthermore, \textbf{\emph{HyperIM\_BRR}} reduces the number of RR sets required to reach the desired approximation guarantee. All the algorithms are implemented in C++. Each algorithm is executed $10$ times to report the average results.

\textbf{\emph{Parameter settings and Evaluation metrics.}} Recall that an error parameter ($\varepsilon$) and a failure probability parameter ($\delta$) are required by each IM algorithm for discovering the seed set. We set $\varepsilon=0.01$ and $\delta=\frac{1}{|V|}$ as the same to the previous work except for evaluating the performances under different error parameters.
We evaluate the IM algorithms from three aspects: (i) influence spread, (ii) the efficiency of RR sets in terms of the number of samples and sampling efficiency, and (iii) the running time of the IM algorithms. Furthermore, we evaluate the four IM algorithms under different settings in terms of cascade models and the error parameters.
\subsection{Influence spread}
\begin{figure*}[hbt!]
\centering
\subfigure[email-Eu-full]{\label{email-Eu-full_INF}
\includegraphics[width=0.18\textwidth,height=0.13\textwidth]{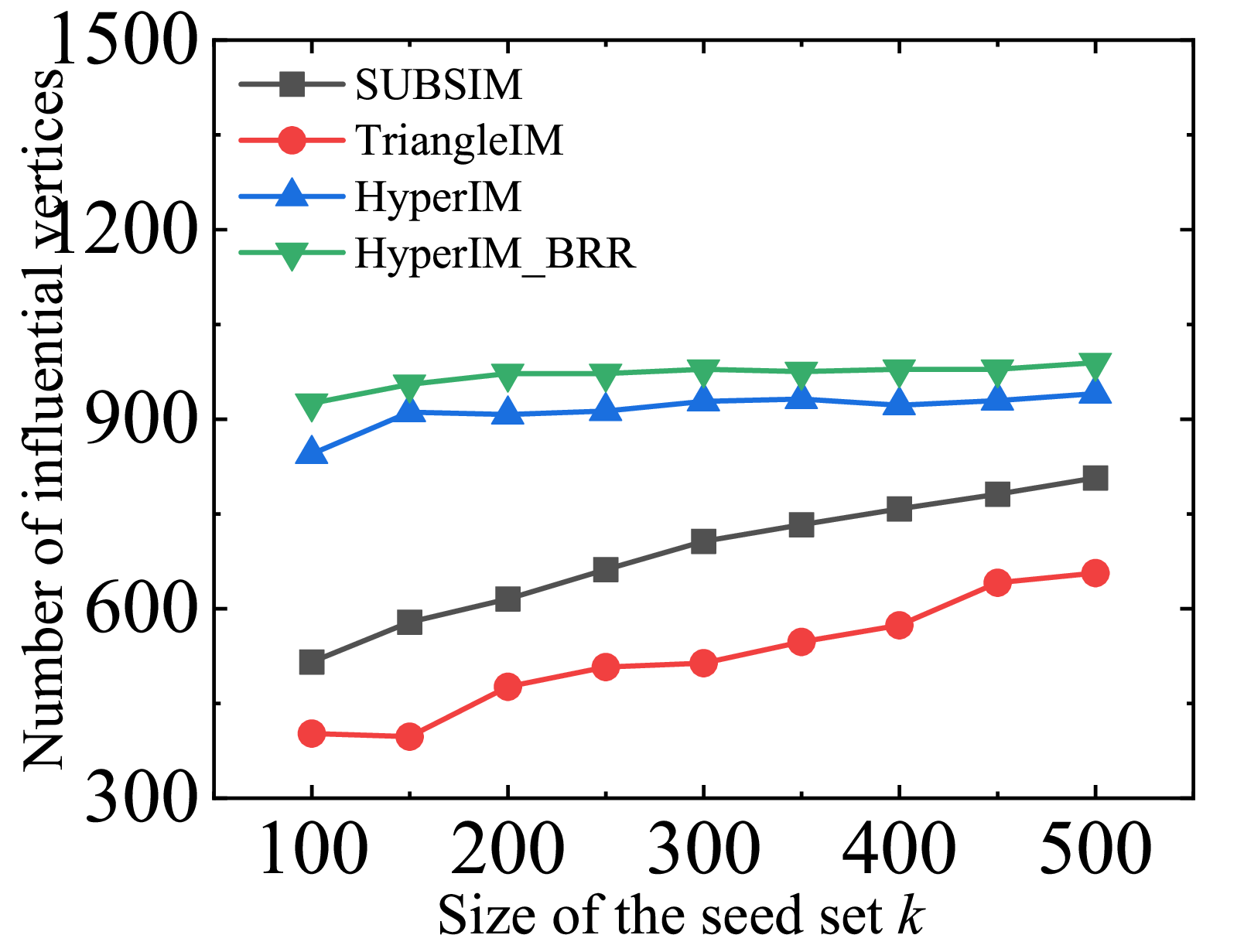}
}\subfigure[tags-ask-ubuntu]{\label{tags-ask-ubuntu_INF}
\includegraphics[width=0.18\textwidth,height=0.13\textwidth]{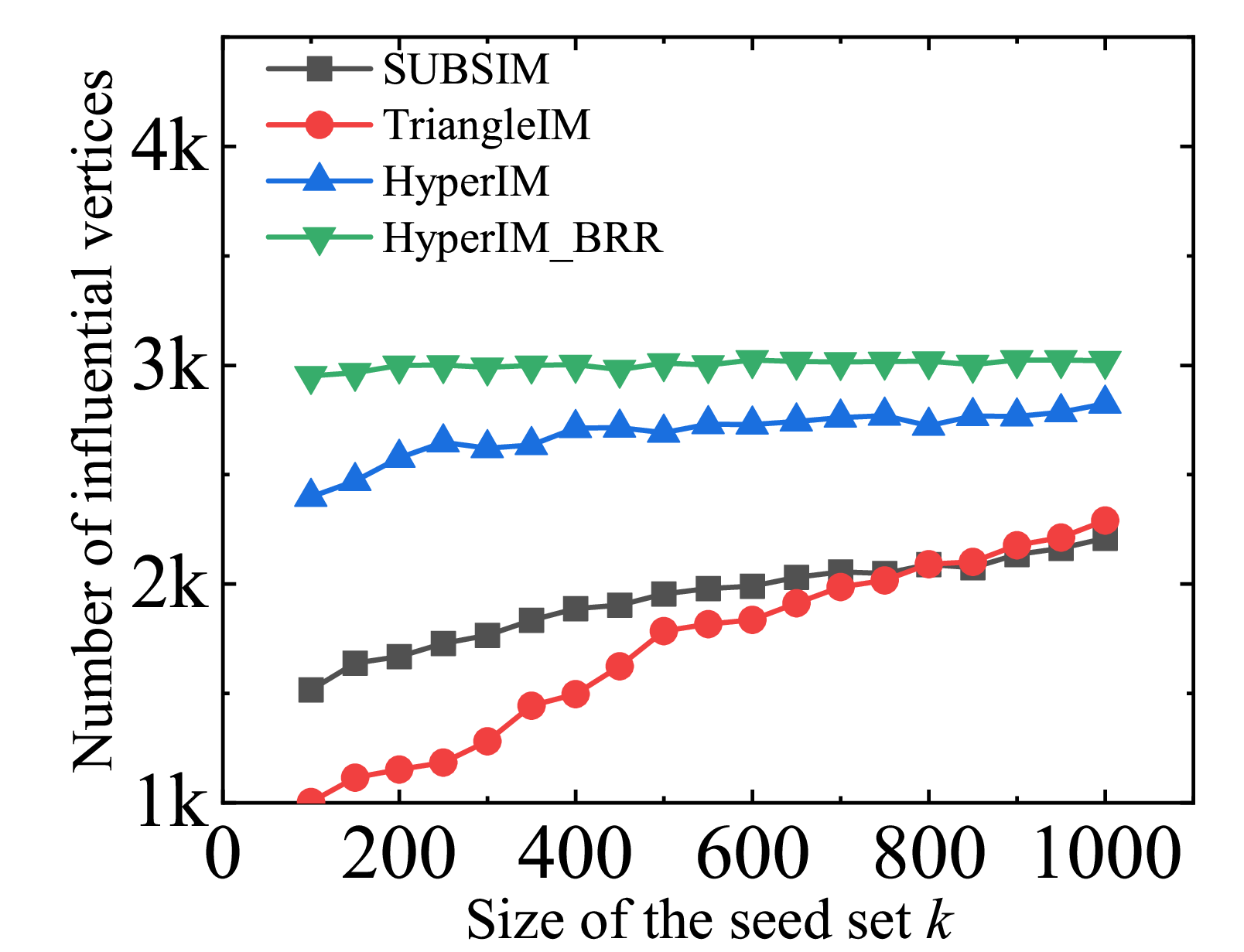}
}\subfigure[NDC-substances-full]{\label{NDC-substances-full_INF}
\includegraphics[width=0.18\textwidth,height=0.13\textwidth]{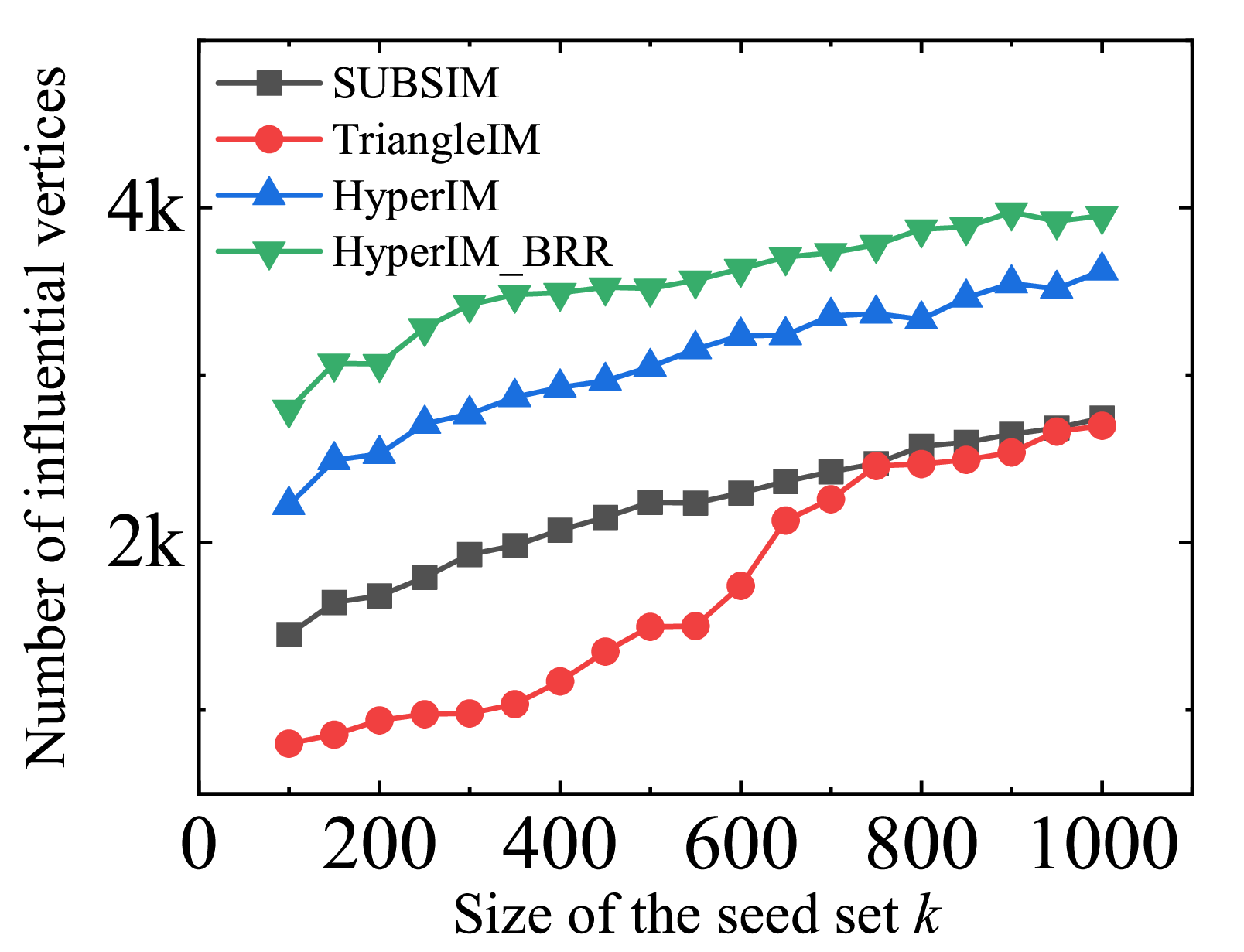}
}\subfigure[threads-ask-ubuntu]{\label{threads-ask-ubuntu_INF}
\includegraphics[width=0.18\textwidth,height=0.13\textwidth]{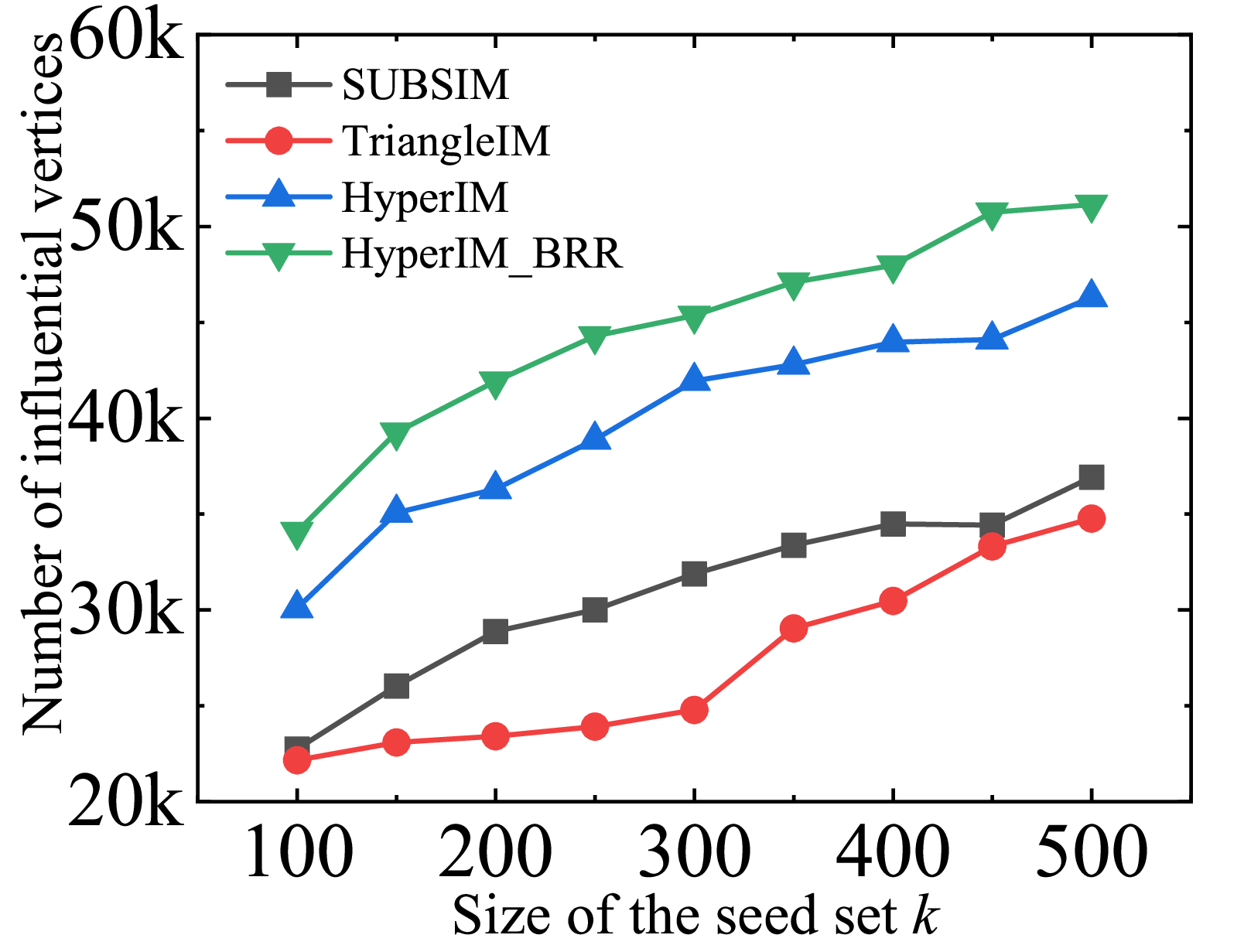}
}\subfigure[coauth-MAG-Geology-full]{\label{coauth-MAG-Geology-full_INF}
\includegraphics[width=0.18\textwidth,height=0.13\textwidth]{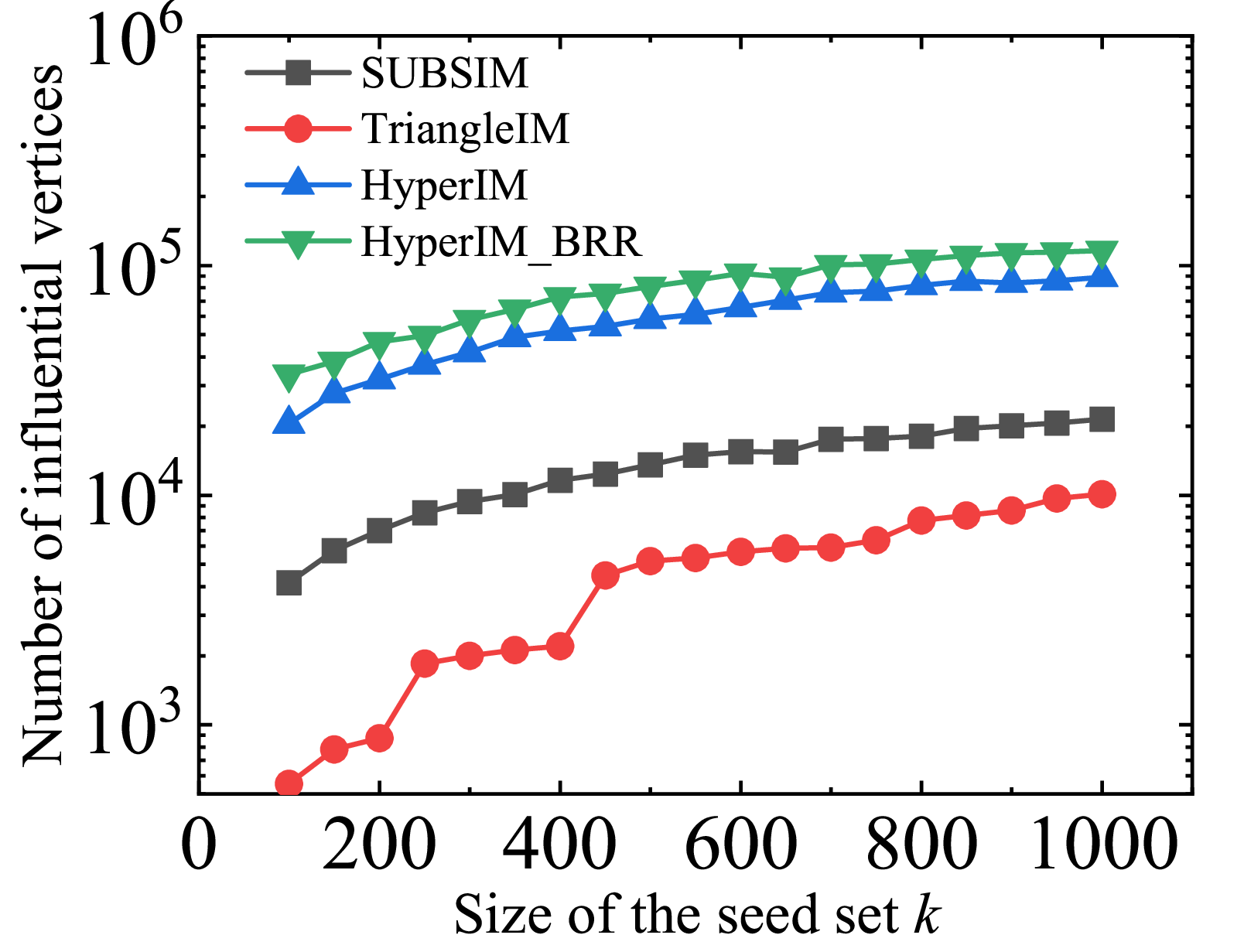}
}\caption{The number of influential vertices with different sizes of seed sets using different IM algorithms under the IC model.}\label{fig_INF_WC}
\end{figure*}

Given a seed set containing $k$ vertices, we evaluate influence spread in terms of the number of the influential vertices.
Figure \ref{fig_INF_WC} shows the expected influences of the seed sets with $k$ ranging from $100$ to $1000$ using the four IM algorithms.
The seed sets of the five hypergraph datasets obtained by \emph{HyperIM} have 1.6X, 1.74X, 1.85X, 1.67X and 4.25X on average higher influence than those obtained by \emph{SUBSIM} over the five datasets. This is because \emph{HyperIM} optimizes the vertex selection by stratified sampling and the higher influential vertices are selected with larger probabilities to be added into the RR sets. Then, the seed set obtained from the higher influential RR sets in \emph{HyperIM} must have the higher influence spread than \emph{SUBSIM} and \emph{TriangleIM}. The seed sets obtained by \emph{TriangleIM} perform the smallest influences and this is because \emph{TriangleIM} tends to discover the vertices from the angle of participate in more triangle formations rather than from the perspective of higher influence to infect the other vertices. Besides inheriting the strengths of \emph{HyperIM} of setting the activation probability, \emph{HyperIM\_BRR} optimizes the generations of RR sets by adjusting the number of vertices from the top layers of the sample sets and thus \emph{HyperIM\_BRR} performs a little better than \emph{SUBSIM} and it has 2.73X on average higher influence spread than \emph{SUBSIM}.
\begin{figure*}[hbt!]
\centering
\subfigure[email-Eu-full]{\label{email-Eu-full_SAM}
\includegraphics[width=0.18\textwidth,height=0.13\textwidth]{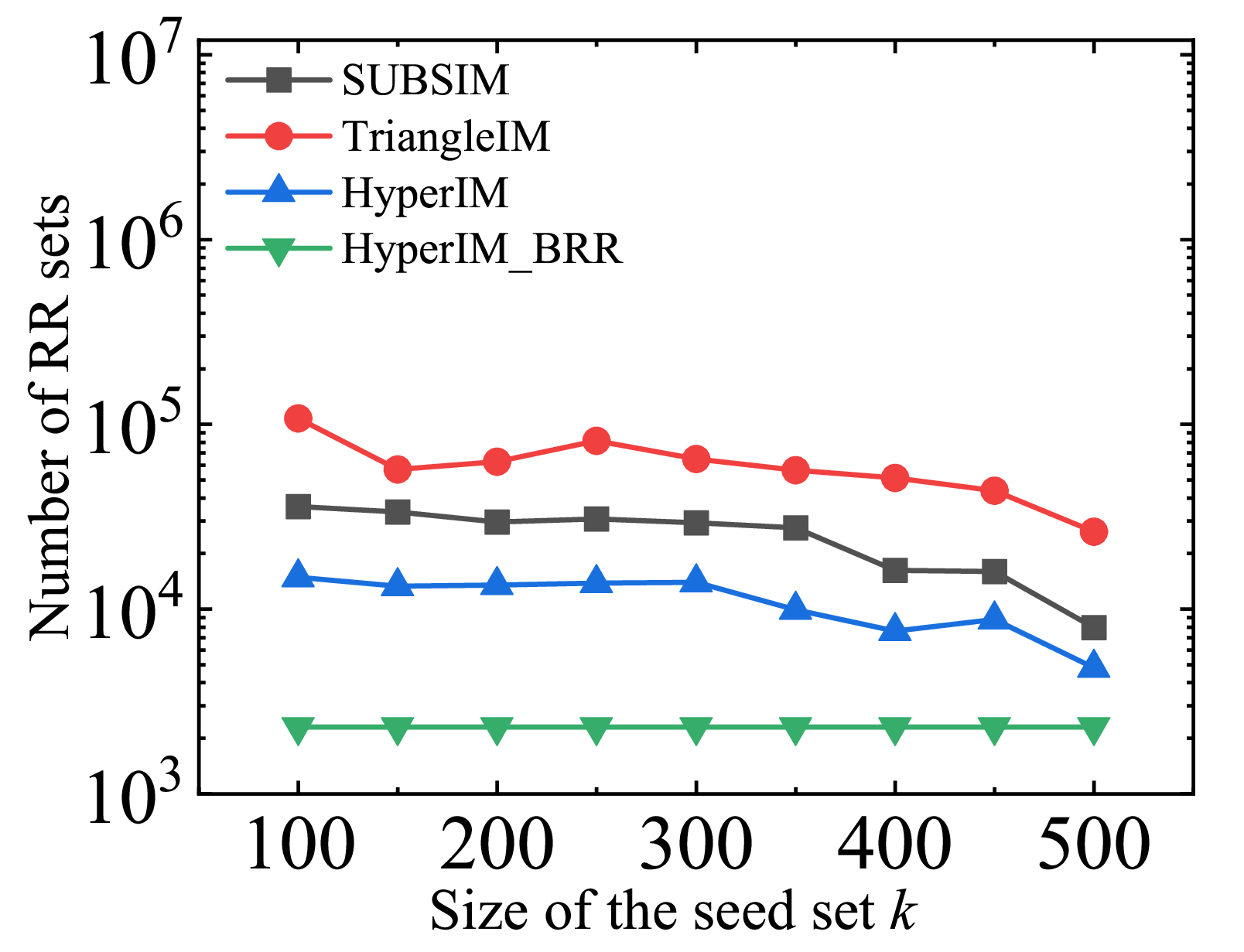}
}\subfigure[tags-ask-ubuntu]{\label{tags-ask-ubuntu_SAM}
\includegraphics[width=0.18\textwidth,height=0.13\textwidth]{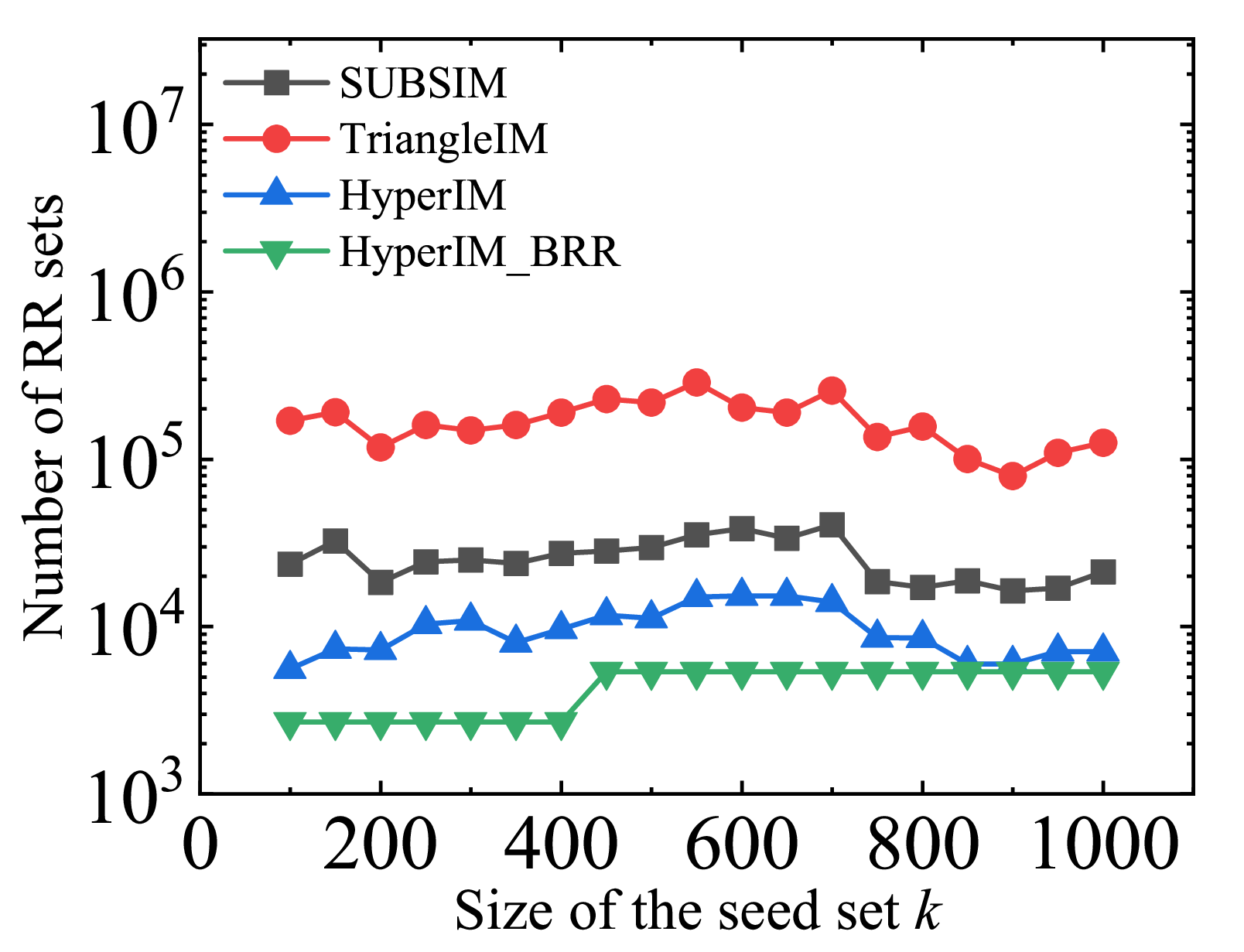}
}\subfigure[NDC-substances-full]{\label{NDC-substances-full_SAM}
\includegraphics[width=0.18\textwidth,height=0.13\textwidth]{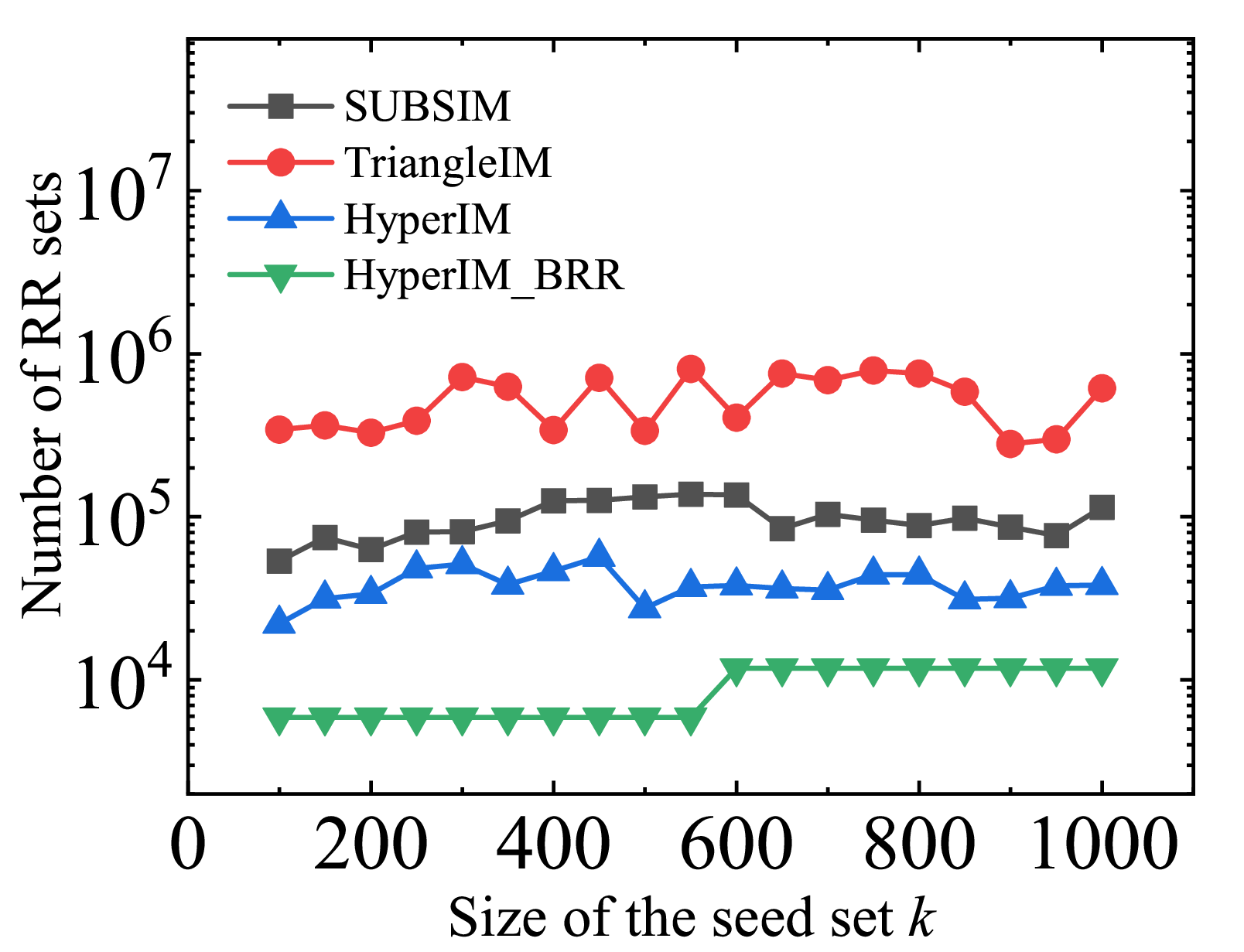}
}\subfigure[threads-ask-ubuntu]{\label{threads-ask-ubuntu_SAM}
\includegraphics[width=0.18\textwidth,height=0.13\textwidth]{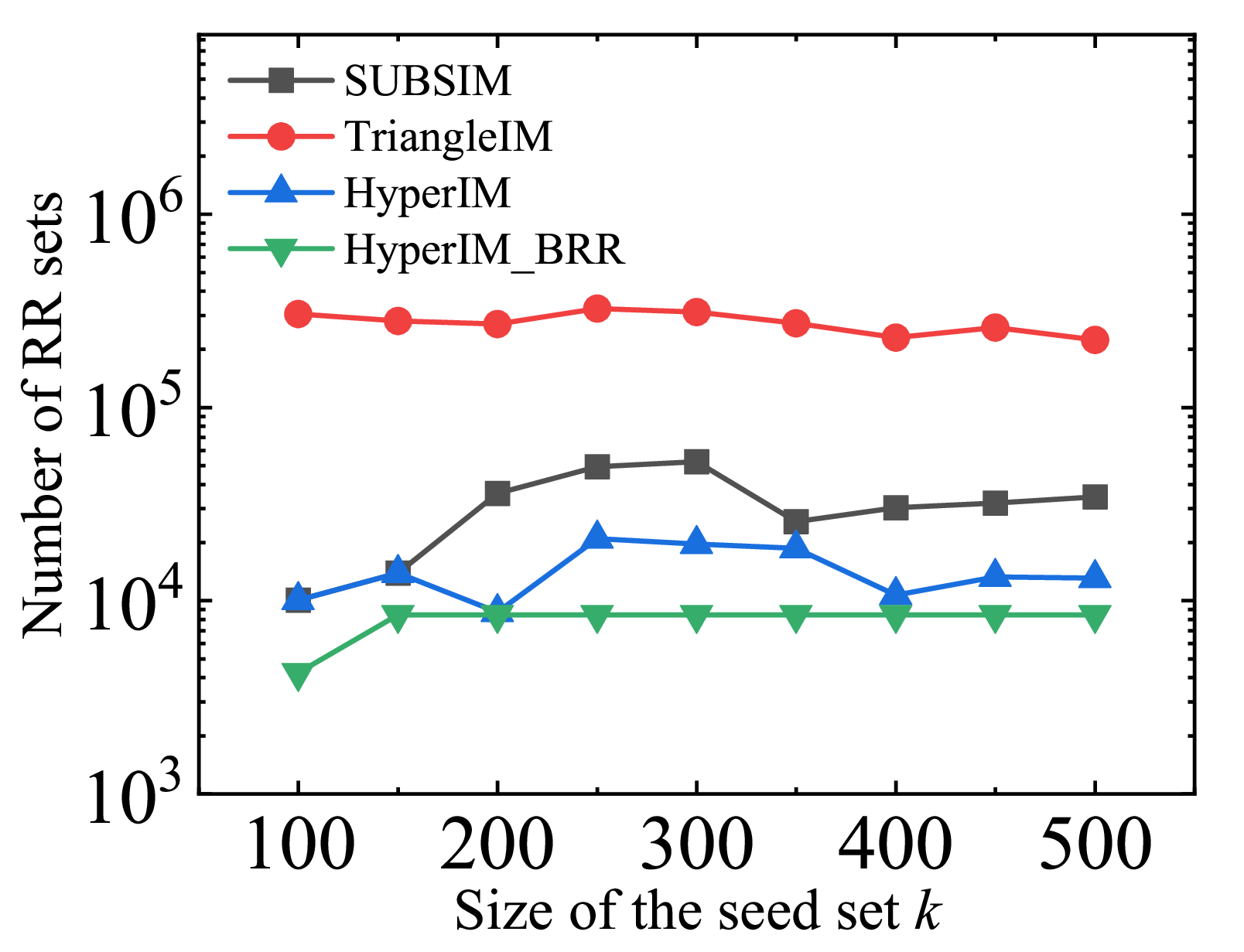}
}\subfigure[coauth-MAG-Geology-full]{\label{coauth-MAG-Geology-full_SAM}
\includegraphics[width=0.18\textwidth,height=0.13\textwidth]{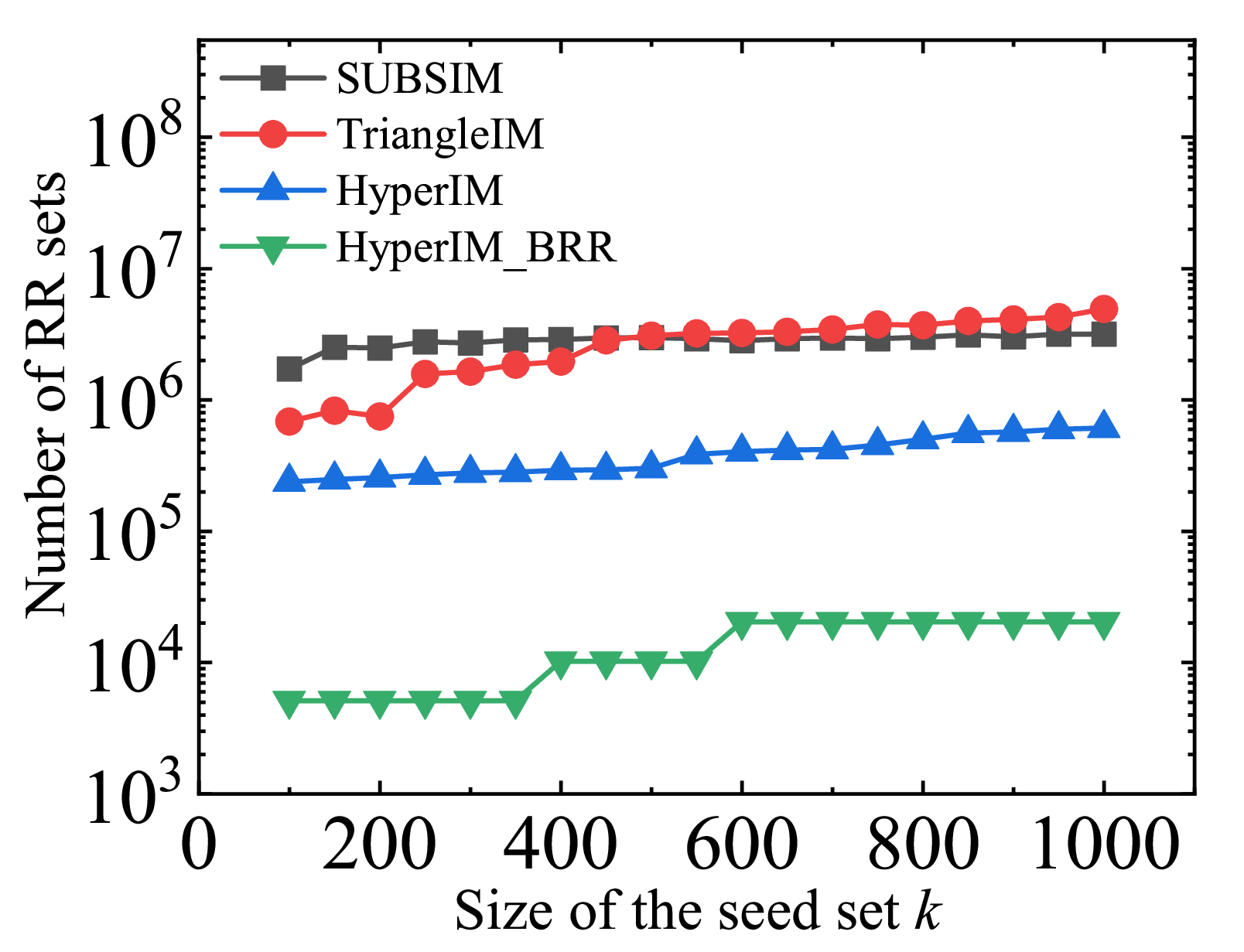}
}\caption{The number of RR sets with different sizes of seed sets using different IM algorithms under the IC model.}\label{fig_sample_WC}
\end{figure*}
\begin{figure}[hbt!]
\centering
\subfigure[Sampling times]{\label{number_efficiency}
\includegraphics[width=0.24\textwidth]{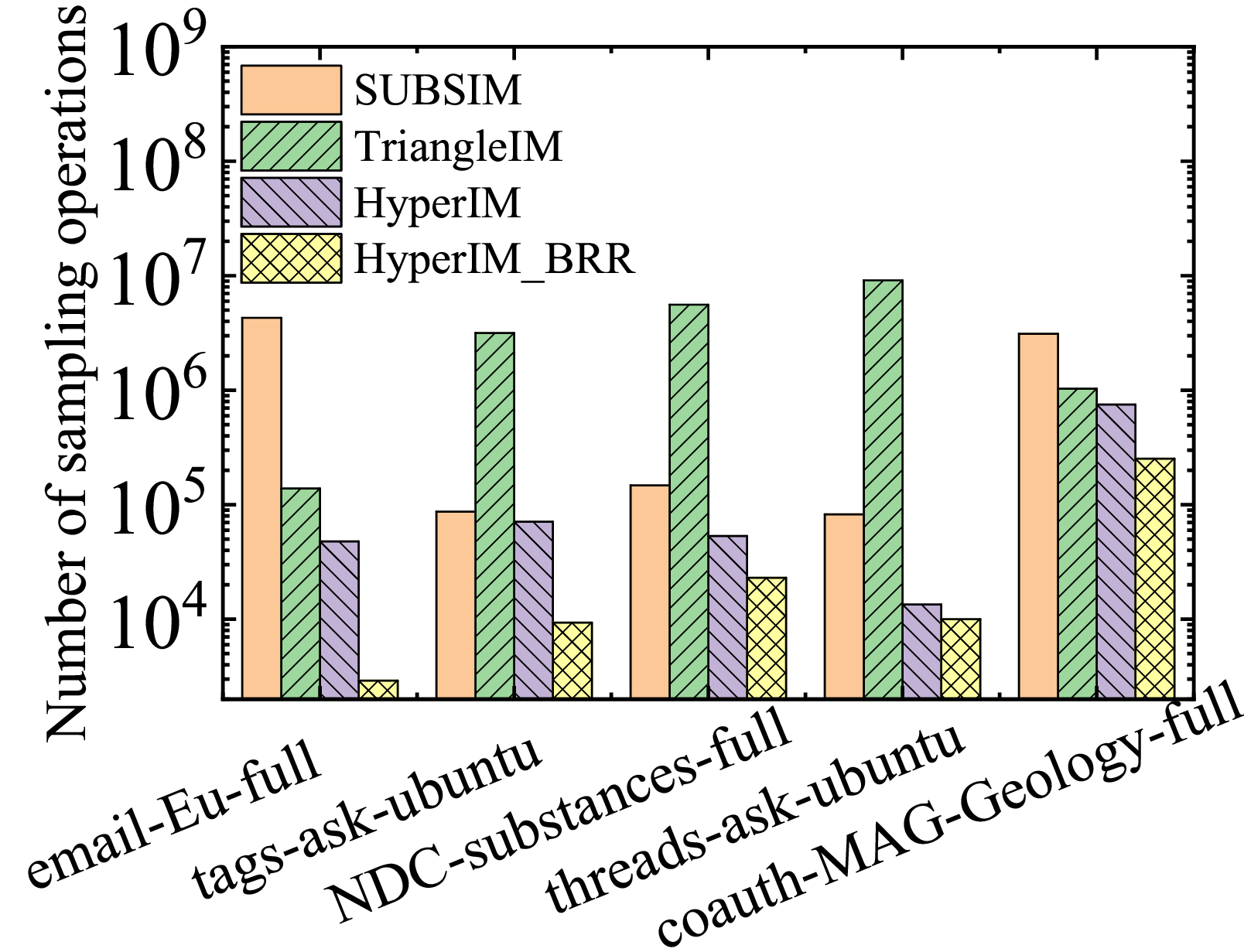}
}\subfigure[Sampling variances]{\label{variance_efficiency}
\includegraphics[width=0.24\textwidth]{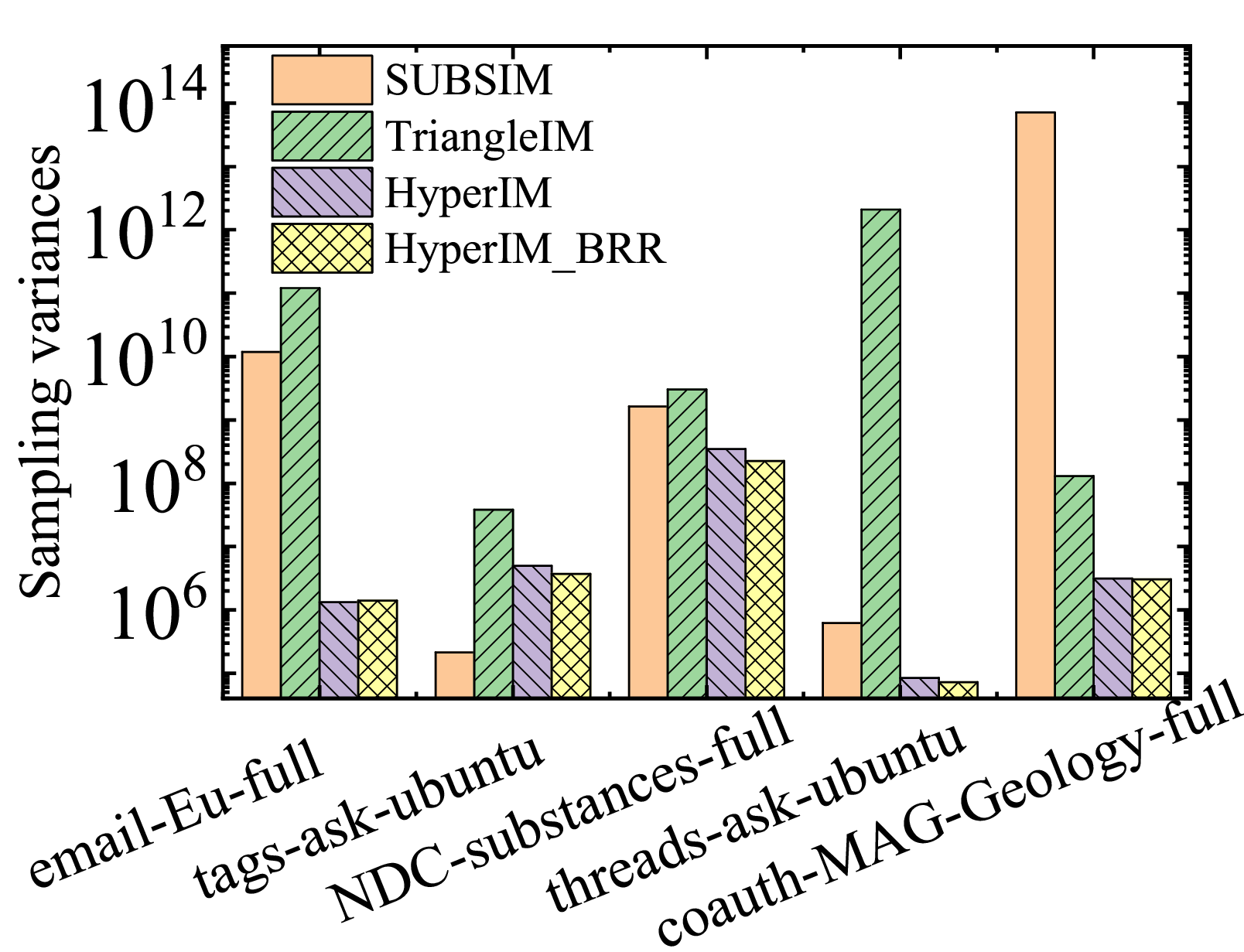}
}\caption{Sampling efficiency.}\label{fig_efficiency}
\end{figure}
\subsection{The efficiency of RR sets}
We evaluate the efficiency of RR sets from two aspects: (i) the number of RR sets required by the IM algorithms to discover the seed sets
and (ii) the number of the sampling operations and the sampling variance to obtain the RR sets.

\emph{The number of RR sets.} As shown in Figure \ref{fig_sample_WC}, with the increase of the size of the seed sets,
\emph{HyperIM\_BRR} and \emph{HyperIM} require the minimum number of RR sets while providing the desired approximate guarantees over the five hypergraphs. Compared to \emph{SUBSIM}, \emph{HyperIM} reduces the number of RR sets by up to at least 1.1X and on average 3X.
In general, \emph{TriangleIM} requires the largest number of RR sets over the datasets because it
needs more RR sets to form the required number of triangles as presented in \cite{hu2023triangular}. Such results confirm the efficiency of stratified sampling combined with the Binomial-based strategy and the Possion-based strategy which are efficient to just require a smaller number of RR sets for obtaining an optimal seed set. Because \emph{HyperIM\_BRR} decreases the upper bound while increasing the lower bound by setting the number of vertices in the top layers, \emph{HyperIM\_BRR} requires smaller numbers of RR sets than \emph{HyperIM}. \emph{HyperIM\_BRR} can thus greatly reduce the number of RR sets by up to 8X compared to \emph{SUBSIM} over the five hypergraph datasets while it can bring in higher influence spread.

\emph{Sampling number and variance.} As shown in Figure \ref{fig_efficiency}(a), \emph{HyperIM\_BRR} and \emph{HyperIM} require two orders of magnitude smaller numbers of sampling operations than \emph{SUBSIM} and \emph{TriangleIM} over the five datasets. Compared to the subset sampling and reverse influence sampling used by \emph{SUBSIM} and \emph{TriangleIM} respectively, when using the same number of sampling operations, the stratified sampling with a Binomial-based strategy and a Possion-based strategy is able to produce a RR set with larger size. Since \emph{SUBSIM} and \emph{TriangleIM} require large numbers of RR sets, they necessitate larger numbers of sampling operations accordingly. Besides the required numbers of samplings, the sampling variance of the total sizes of the RR sets is an important to evaluate the efficiency of the IM algorithms. Figure \ref{fig_efficiency}(b) shows that \emph{HyperIM\_BRR} and \emph{HyperIM} have much smaller sampling variances than the existing two IM algorithms. The distributions of sampling operations decide naturally the sampling variances such that the Geometric-based distribution followed by \emph{SUBSIM} leads to larger sampling variance than the Binomial-based distribution or the Possion-based distribution when generating the RR sets from the same sample sets. The distribution of sampling operations in \emph{TriangleIM} follows \emph{(0-1)}-distribution which causes larger sampling variances than that of the Binomial-based distribution and exhibits the same sampling variances as that of the Possion-based distribution. However, \emph{TriangleIM} requires to execute a large number of sampling operations to obtain the RR sets that results in large sampling variances. Since \emph{HyperIM\_BRR} reduces the number of required RR sets,  it brings in a smaller number of sampling operations and exhibits smaller sampling variance than \emph{HyperIM}.
\begin{figure*}[hbt!]
\centering
\subfigure[email-Eu-full]{\label{email-Eu-full_time}
\includegraphics[width=0.18\textwidth,height=0.13\textwidth]{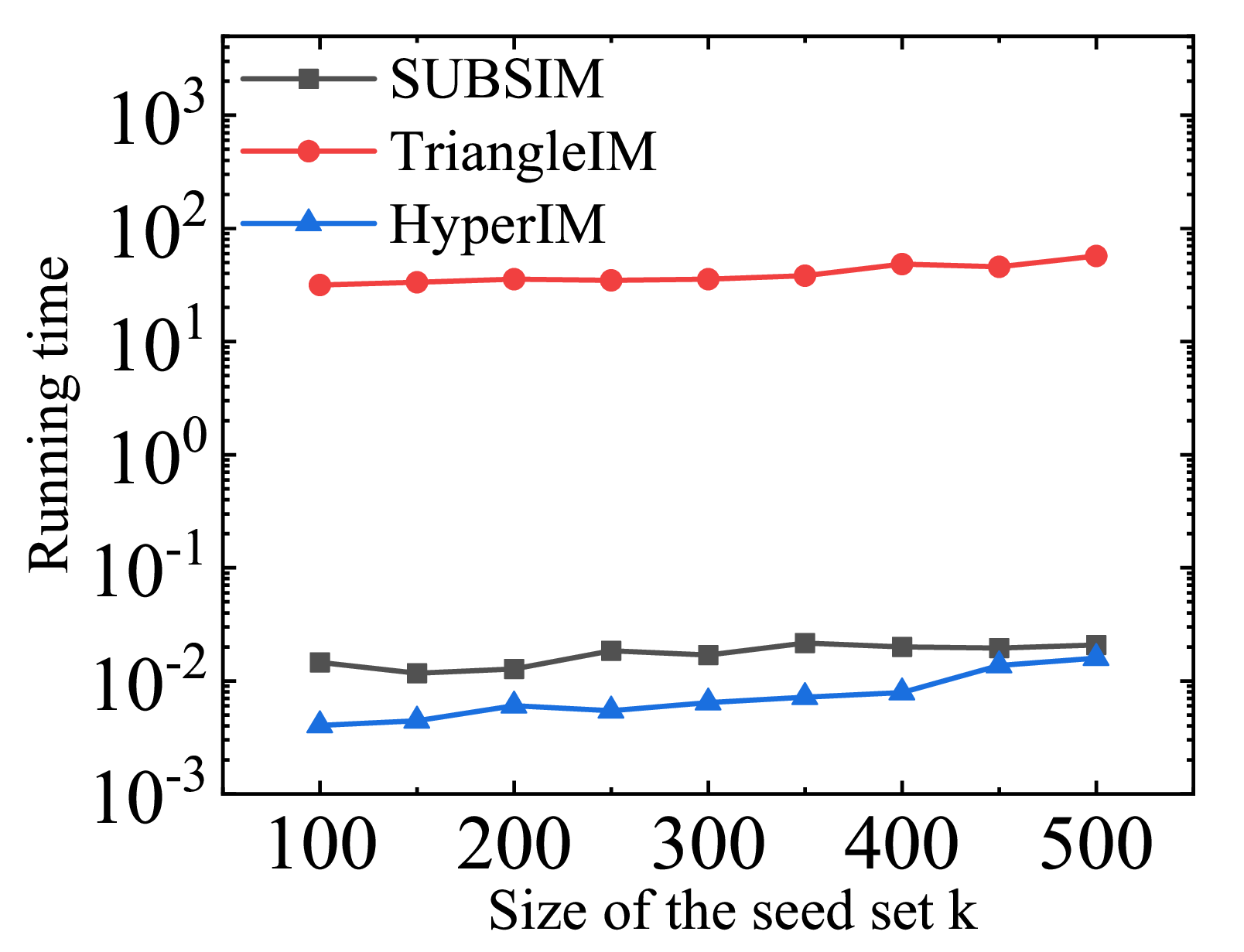}
}\subfigure[tags-ask-ubuntu]{\label{tags-ask-ubuntu_time}
\includegraphics[width=0.18\textwidth,height=0.13\textwidth]{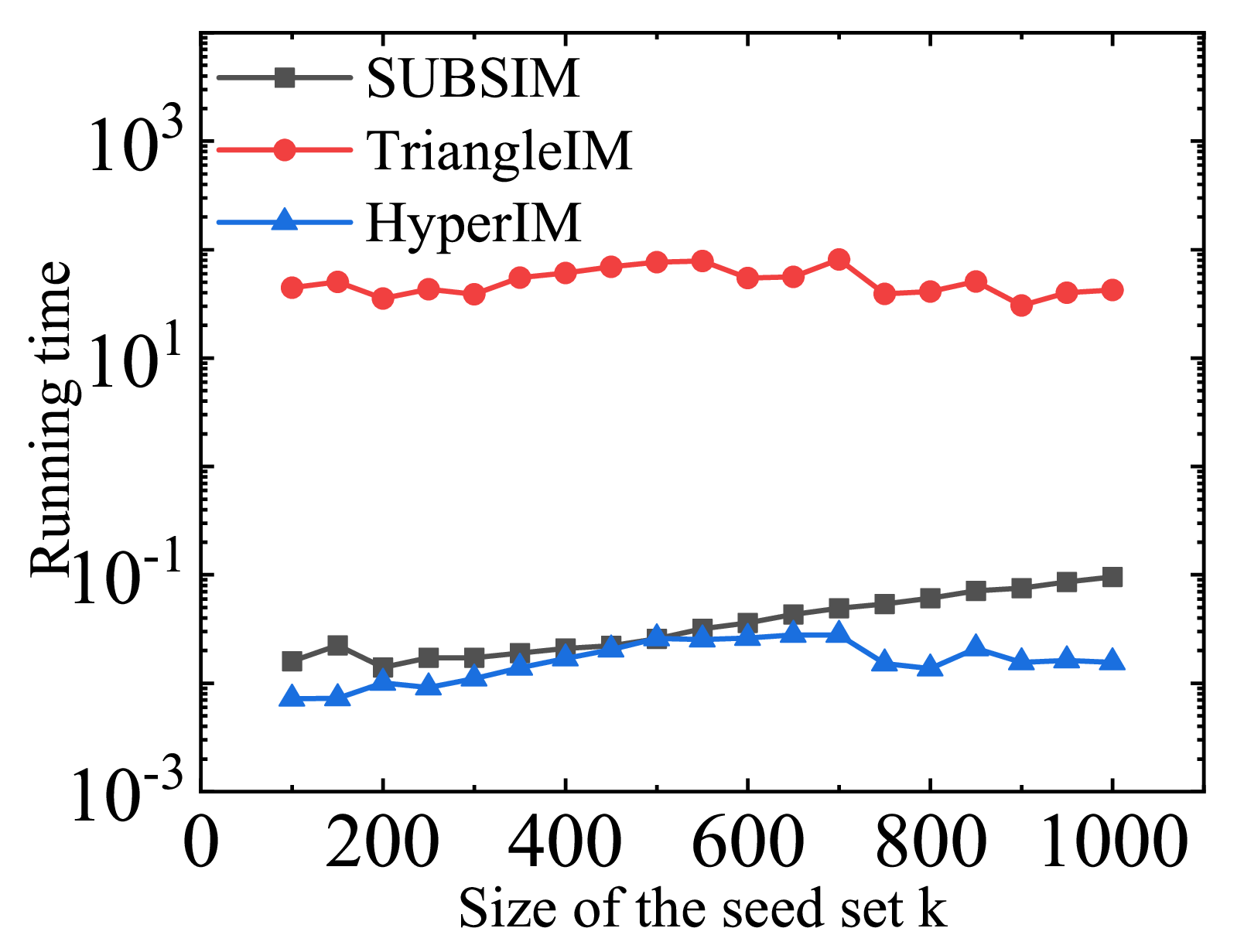}
}\subfigure[NDC-substances-full]{\label{NDC-substances-full_time}
\includegraphics[width=0.18\textwidth,height=0.13\textwidth]{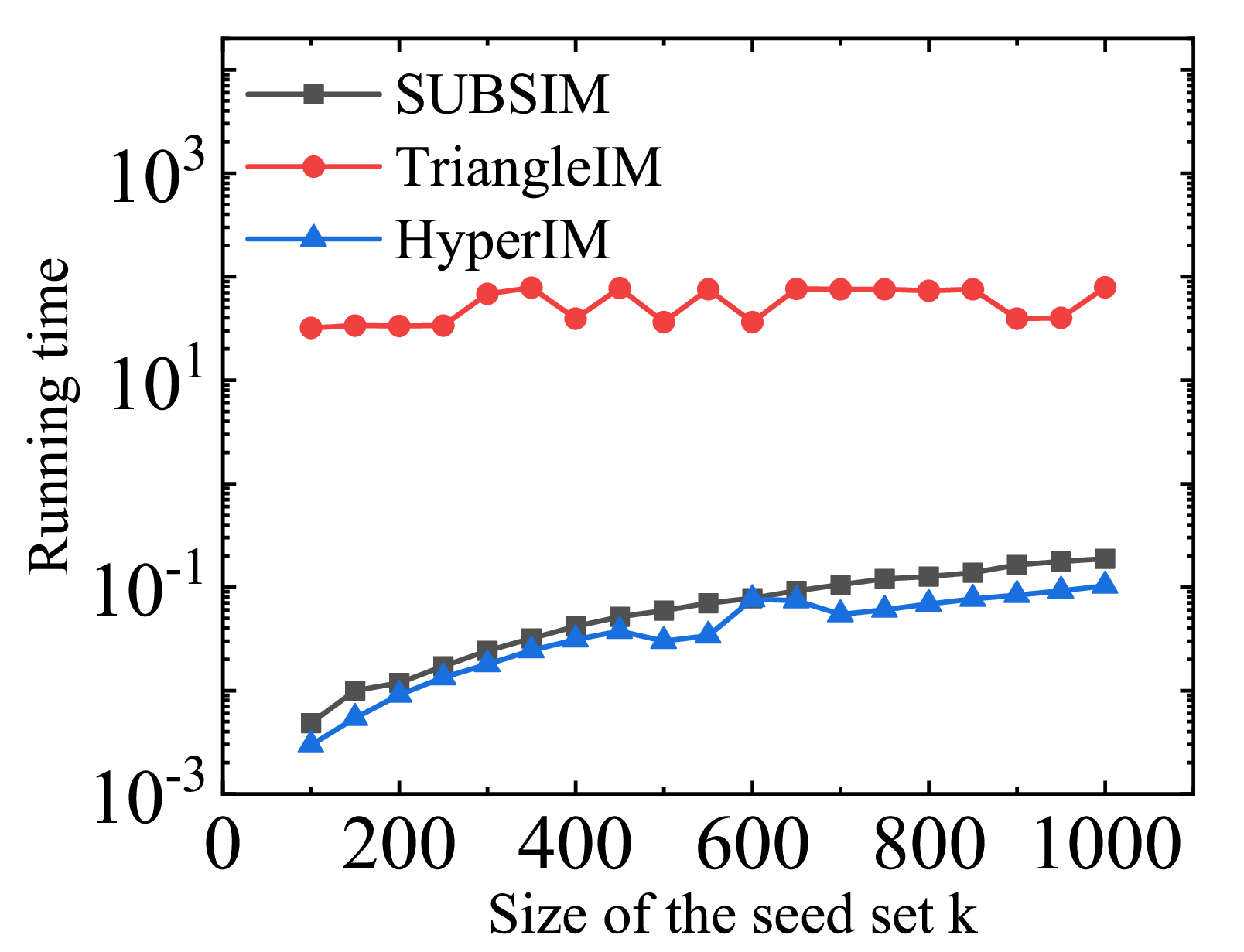}
}\subfigure[threads-ask-ubuntu]{\label{threads-ask-ubuntu_time}
\includegraphics[width=0.18\textwidth,height=0.13\textwidth]{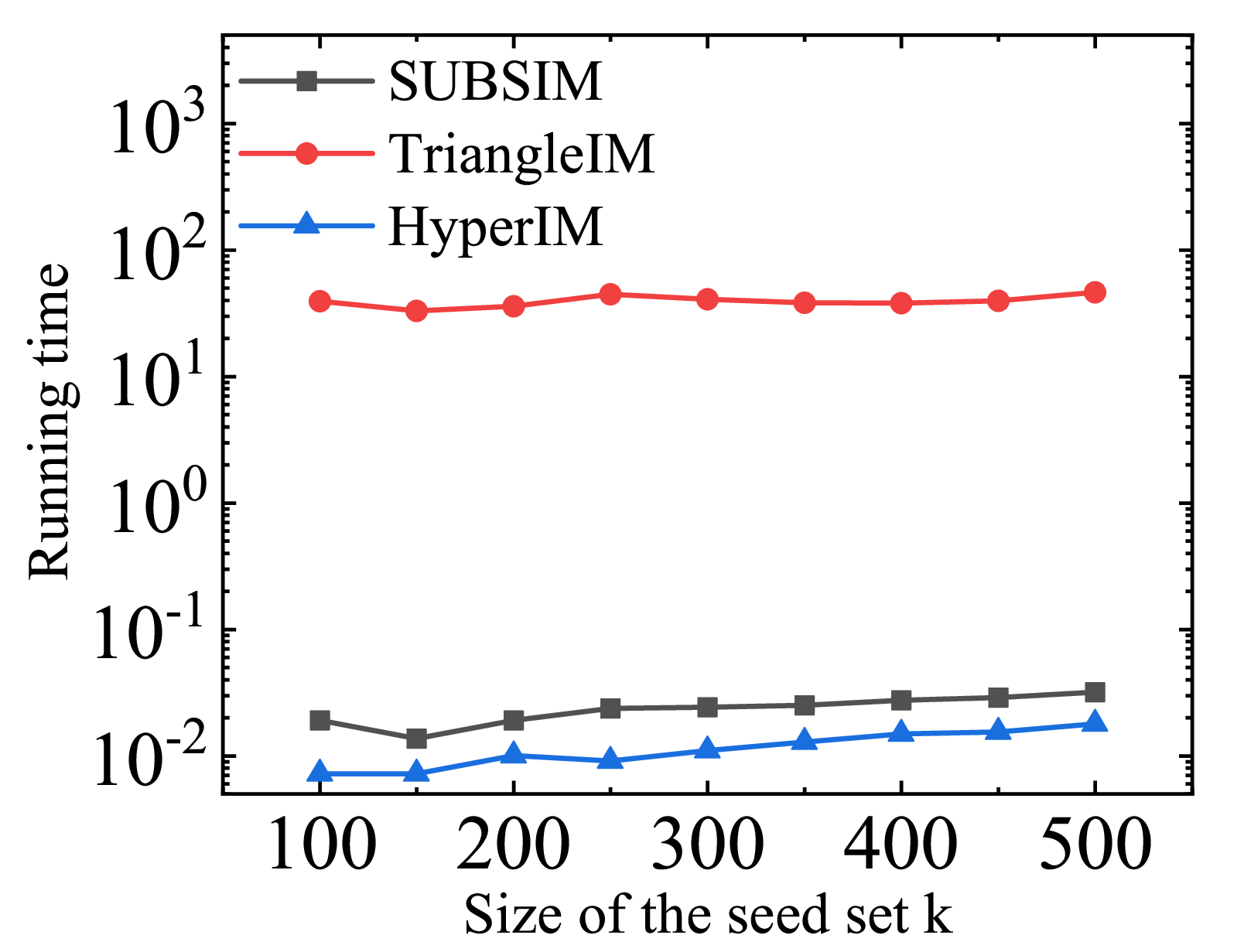}
}\subfigure[coauth-MAG-Geology-full]{\label{coauth-MAG-Geology-full_time}
\includegraphics[width=0.18\textwidth,height=0.13\textwidth]{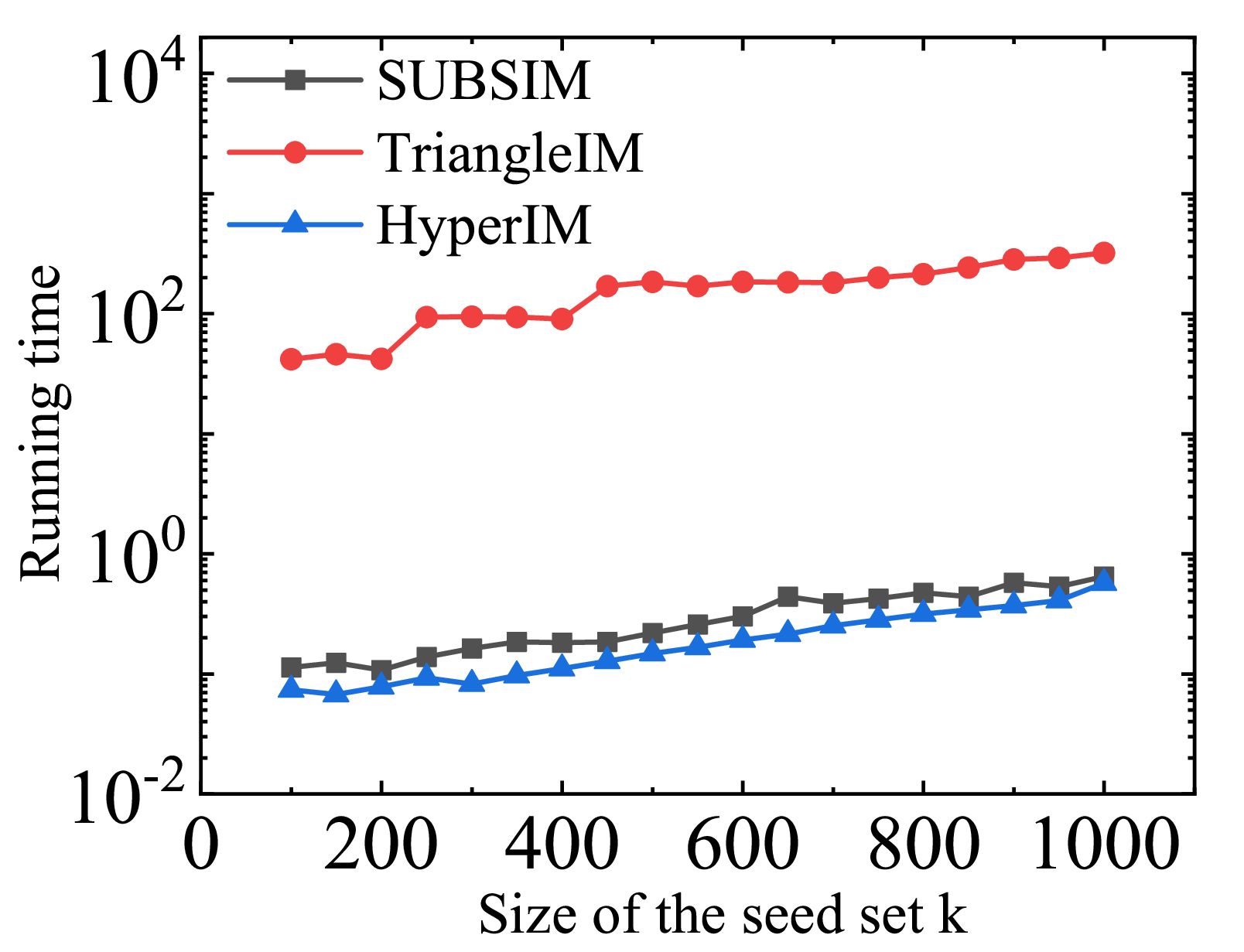}
}\caption{Running time with different sizes of seed sets using different IM algorithms under the IC model.}\label{fig_time}
\end{figure*}
\begin{figure*}[hbt!]
\centering
\subfigure[email-Eu-full]{\label{email-Eu-full_SAMvaryIC}
\includegraphics[width=0.18\textwidth,height=0.13\textwidth]{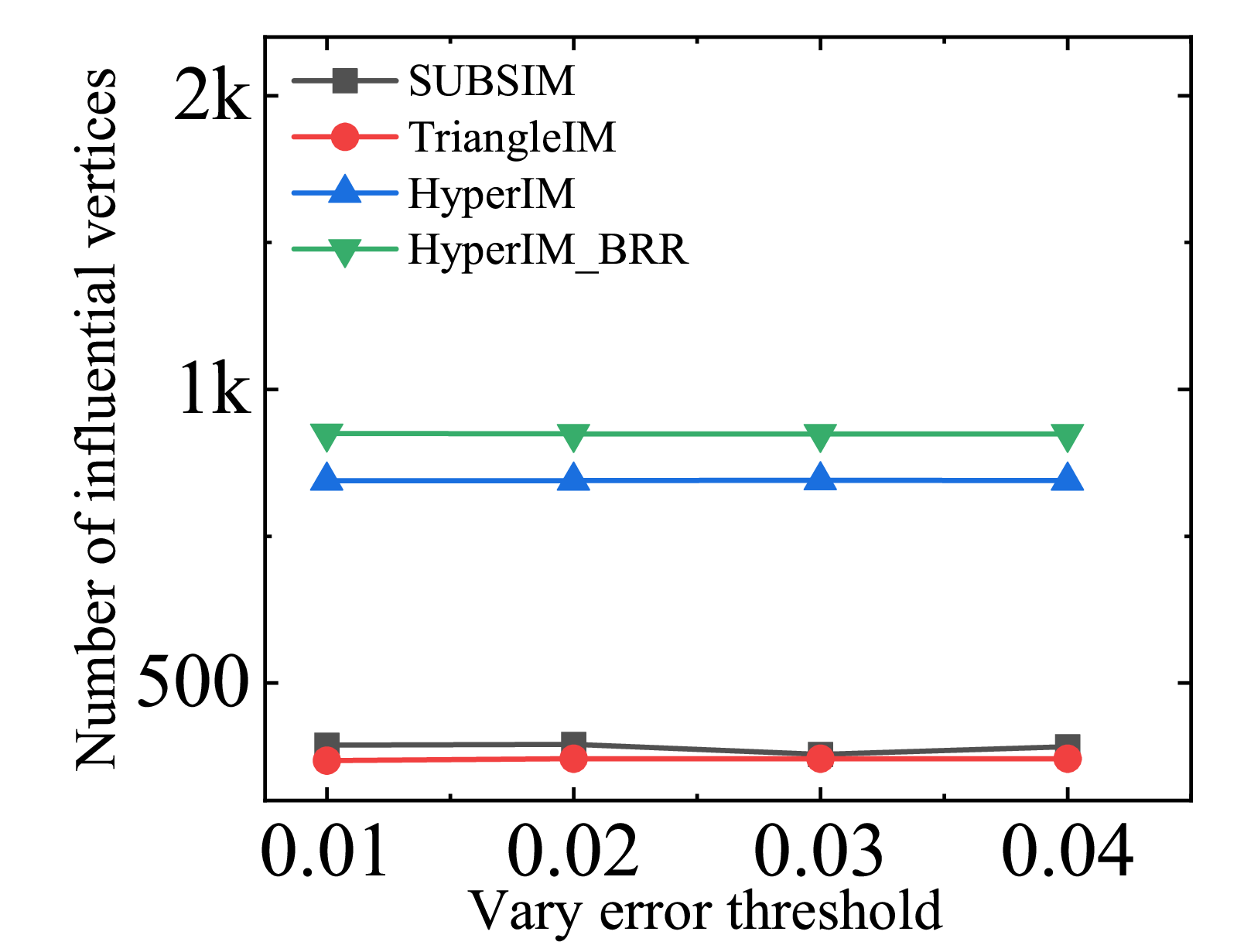}
}\subfigure[tags-ask-ubuntu]{\label{tags-ask-ubuntu_SAMvaryIC}
\includegraphics[width=0.18\textwidth,height=0.13\textwidth]{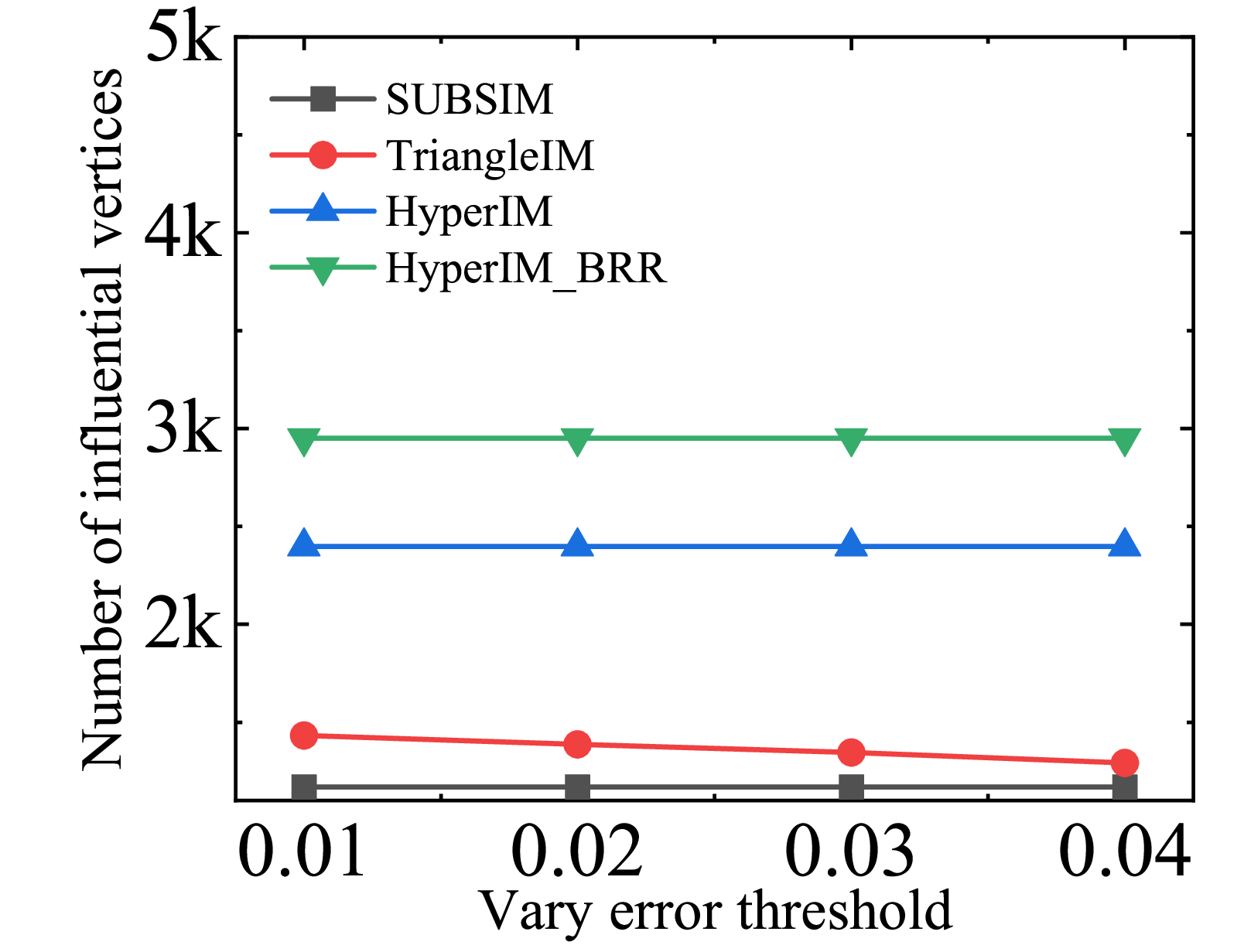}
}\subfigure[NDC-substances-full]{\label{NDC-substances-full_SAMvaryIC}
\includegraphics[width=0.18\textwidth,height=0.13\textwidth]{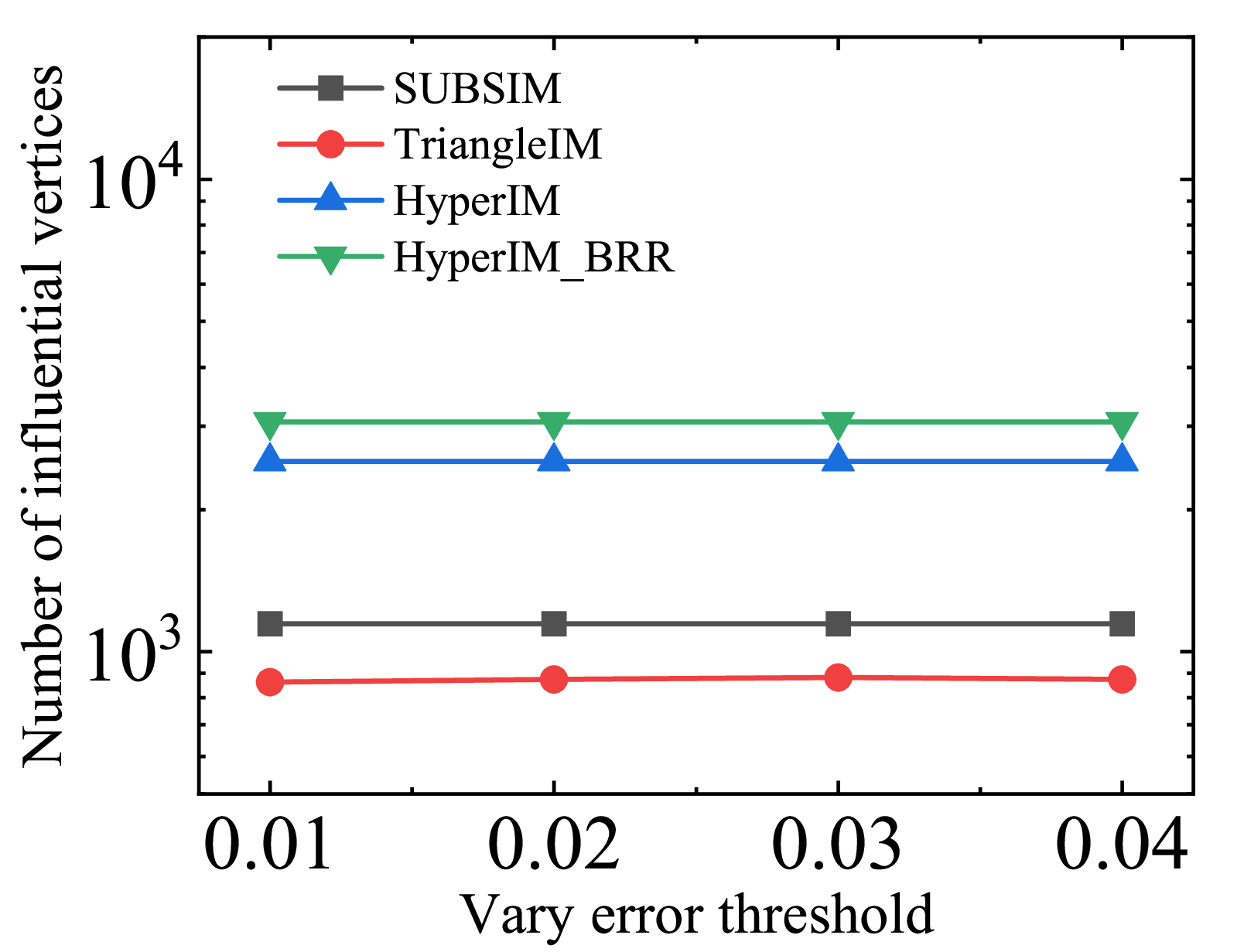}
}\subfigure[threads-ask-ubuntu]{\label{threads-ask-ubuntu_SAMIC}
\includegraphics[width=0.18\textwidth,height=0.13\textwidth]{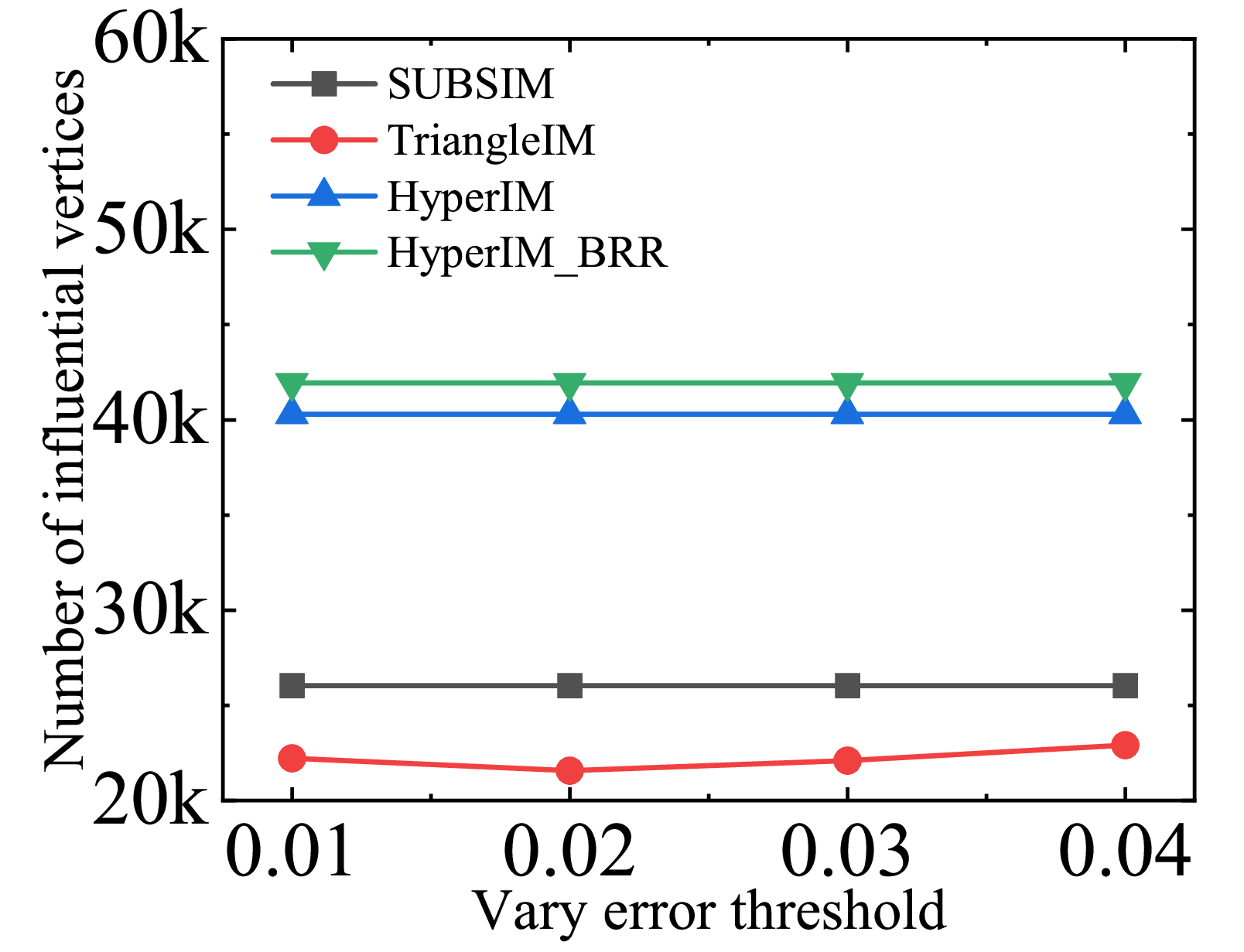}
}\subfigure[coauth-MAG-Geology-full]{\label{coauth-MAG-Geology-full_SAMvaryIC}
\includegraphics[width=0.18\textwidth,height=0.13\textwidth]{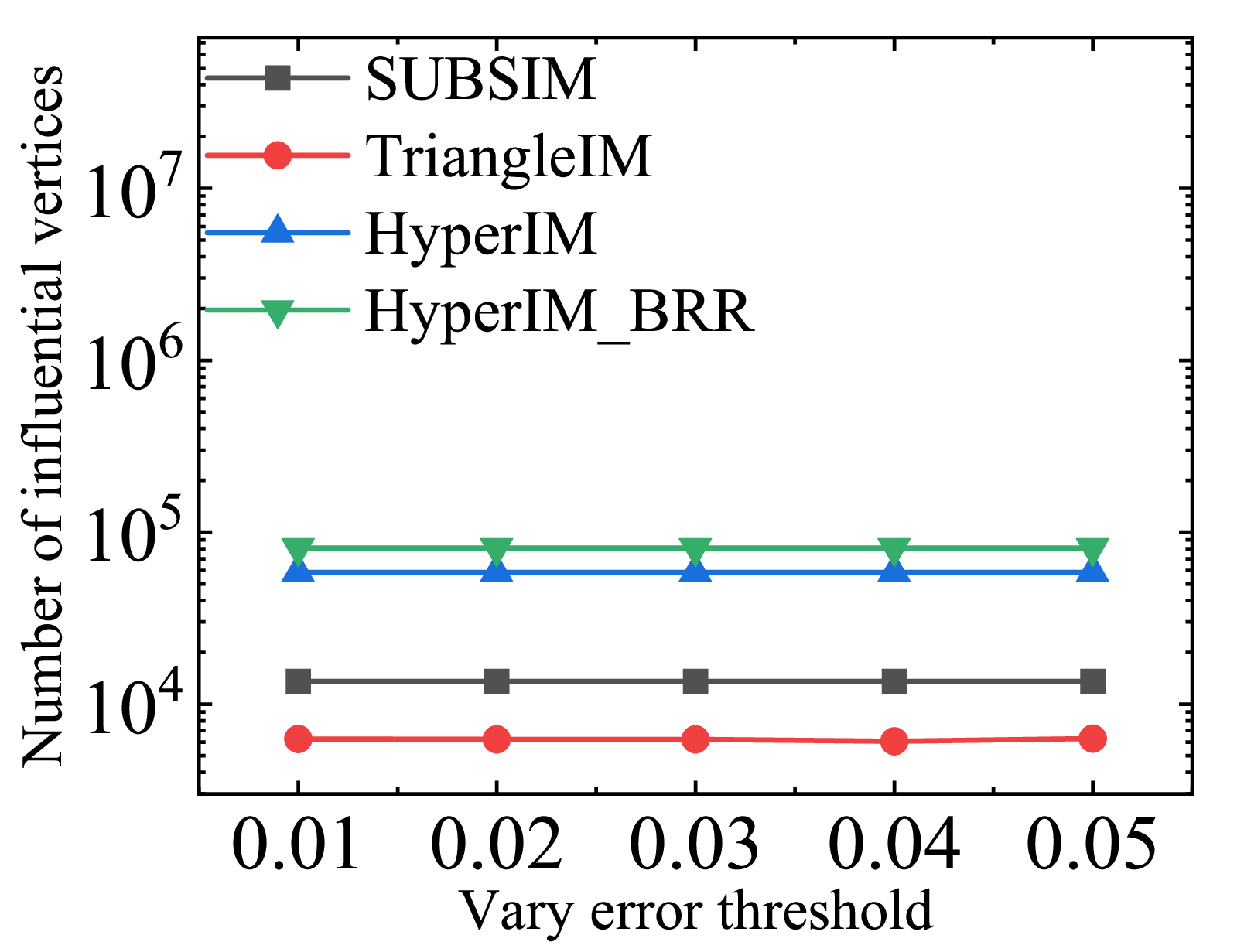}
}\caption{The number of influence vertices with vary thresholds under the LT model.}\label{fig_sample_SAMvaryIC}
\end{figure*}
\begin{figure*}[hbt!]
\centering
\subfigure[email-Eu-full]{\label{email-Eu-full_SAMLT}
\includegraphics[width=0.18\textwidth,height=0.13\textwidth]{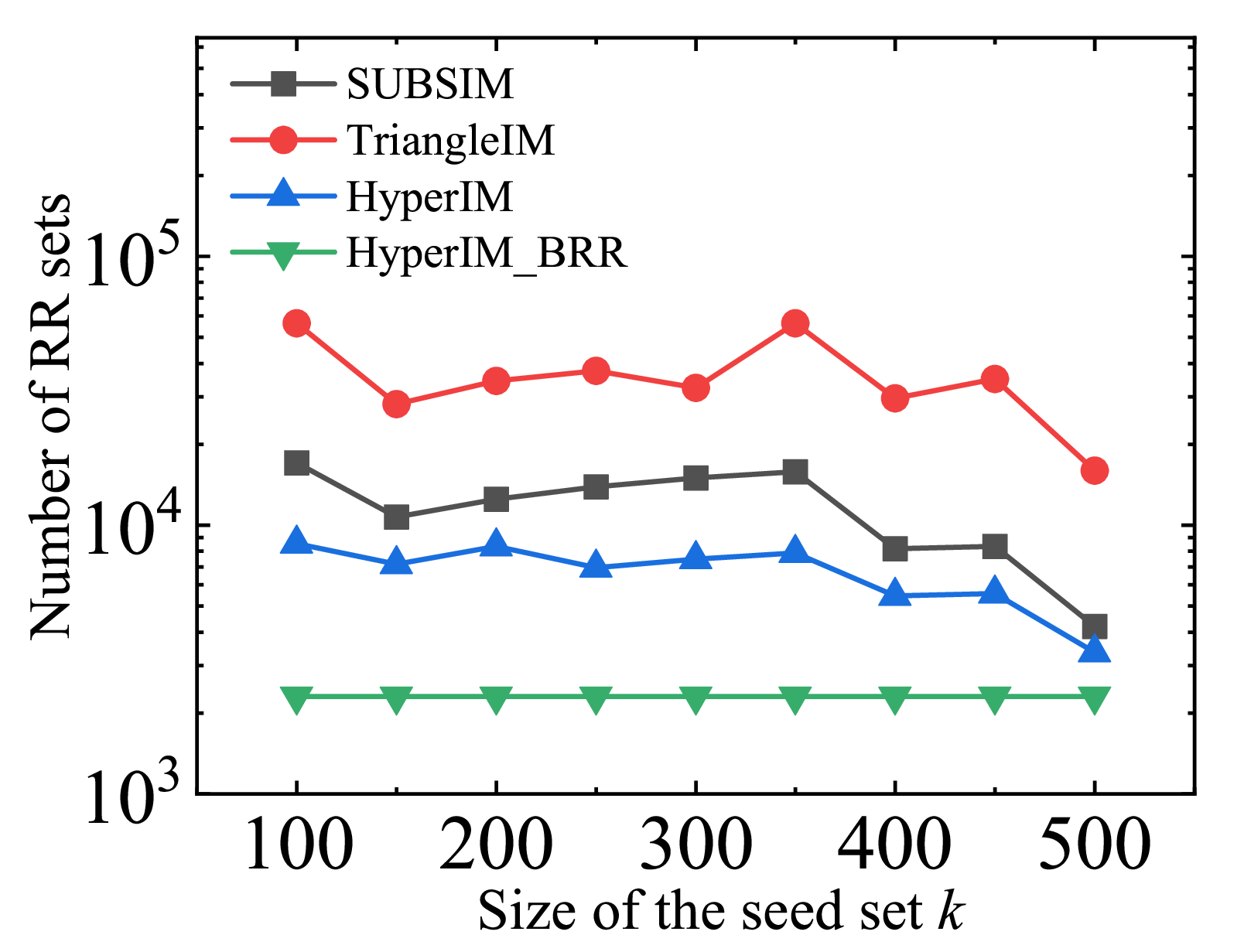}
}\subfigure[tags-ask-ubuntu]{\label{tags-ask-ubuntu_SAMIC}
\includegraphics[width=0.18\textwidth,height=0.13\textwidth]{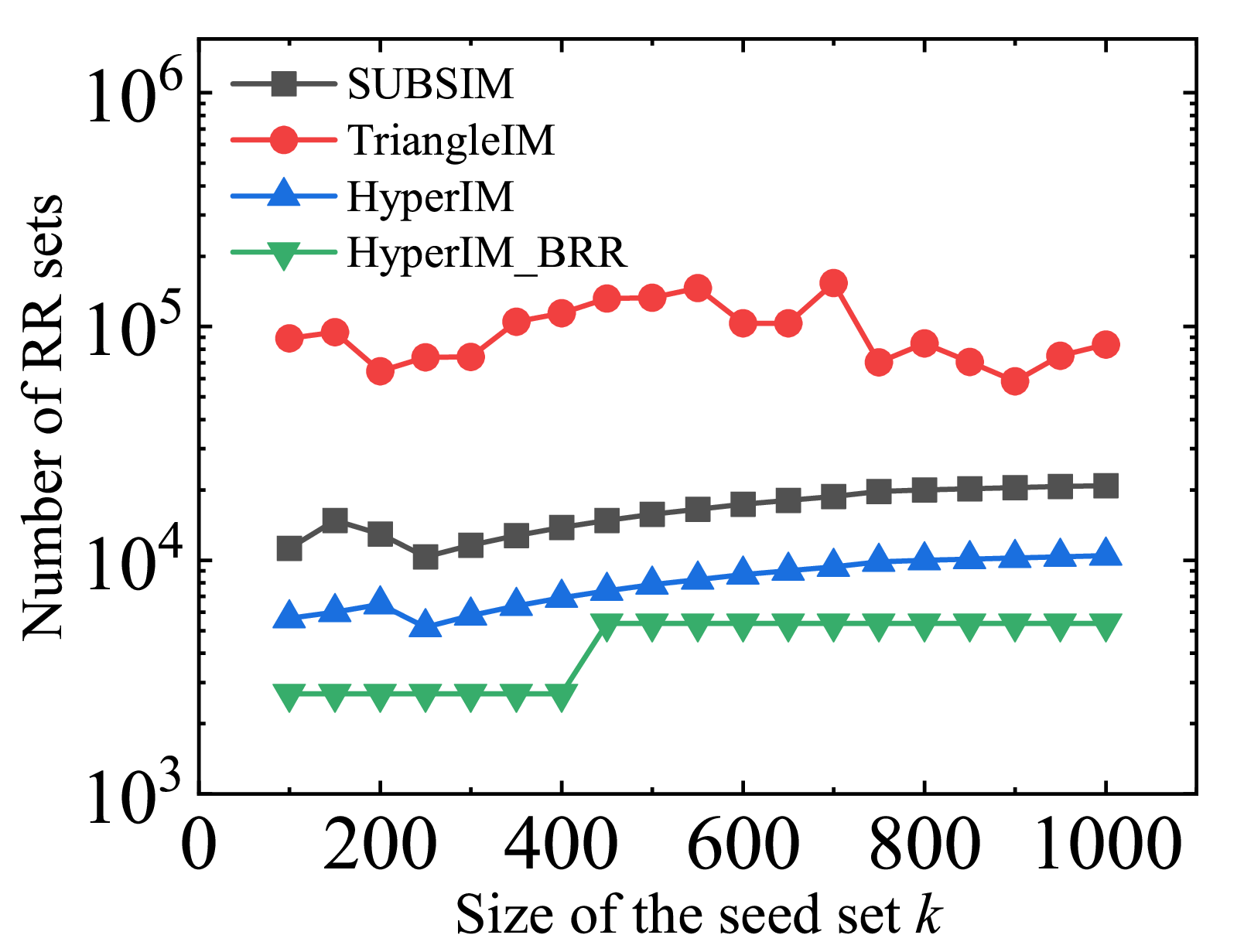}
}\subfigure[NDC-substances-full]{\label{NDC-substances-full_SAMIC}
\includegraphics[width=0.18\textwidth,height=0.13\textwidth]{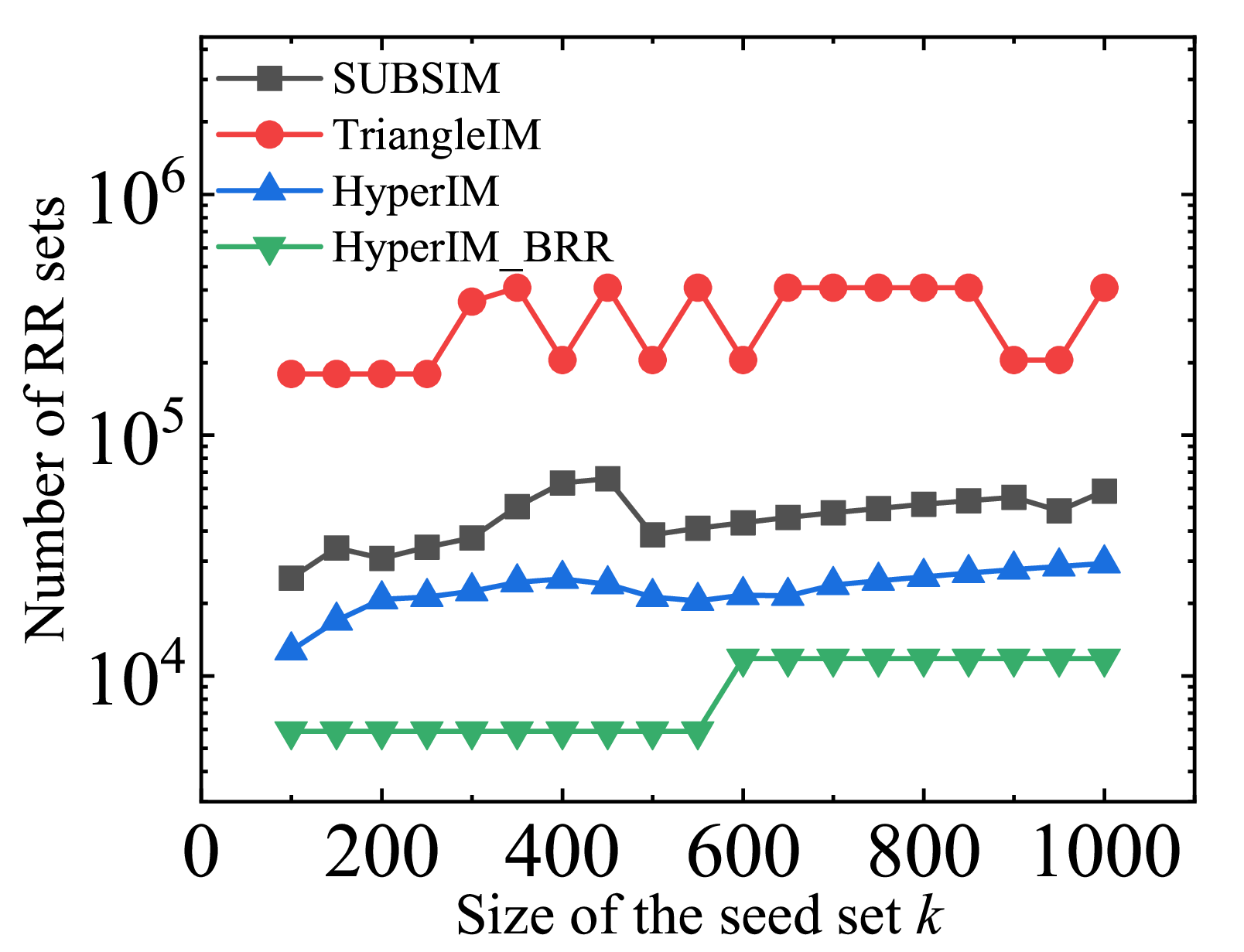}
}\subfigure[threads-ask-ubuntu]{\label{threads-ask-ubuntu_SAMIC}
\includegraphics[width=0.18\textwidth,height=0.13\textwidth]{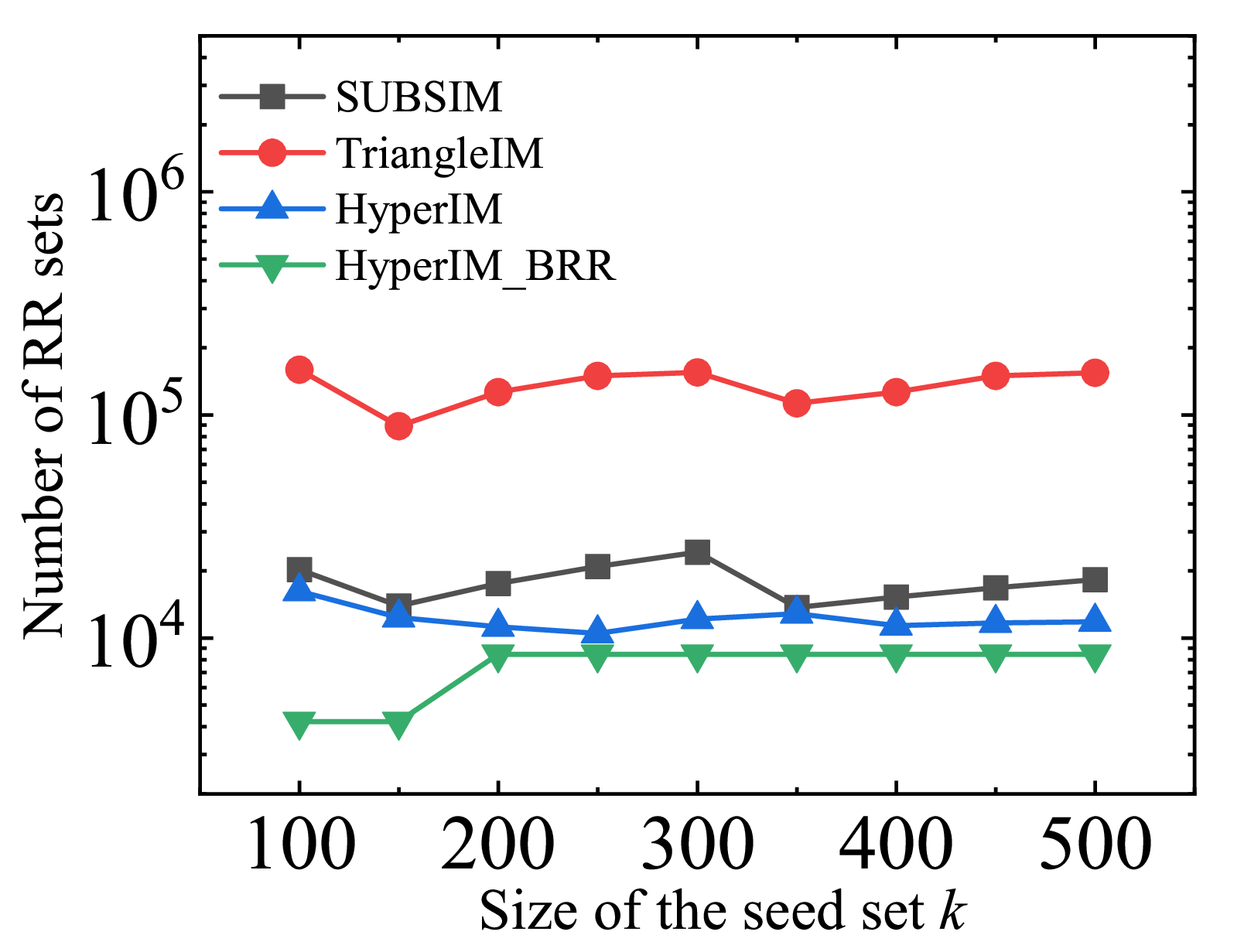}
}\subfigure[coauth-MAG-Geology-full]{\label{coauth-MAG-Geology-full_SAMIC}
\includegraphics[width=0.18\textwidth,height=0.13\textwidth]{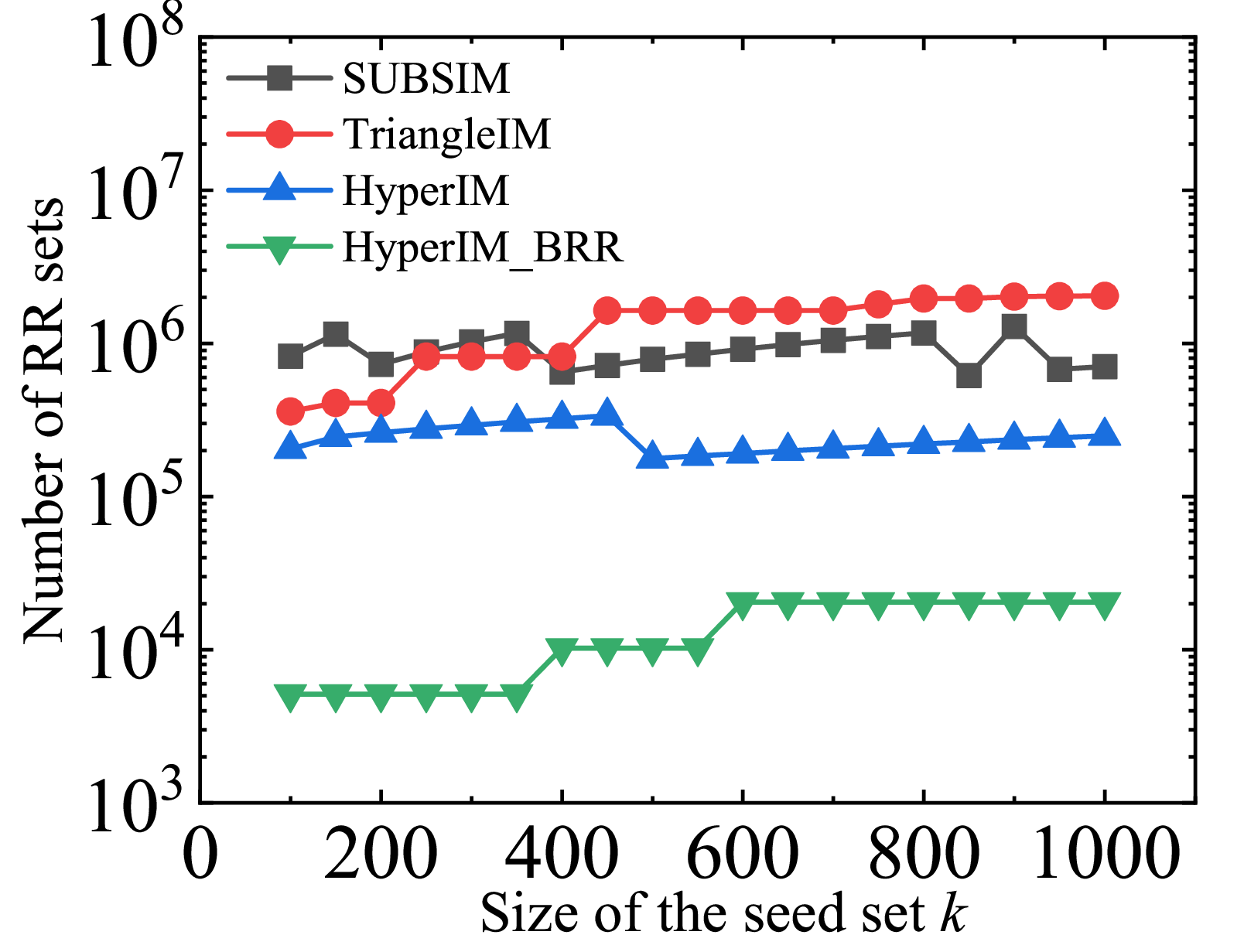}
}\caption{The number of RR sets using different IM algorithms under the LT model.}\label{fig_sample_RR_LT}
\end{figure*}
\subsection{Running time}
Figure \ref{fig_time} shows that \emph{HyperIM} exhibits at least 60\%, at most 10.7X and 3.18X on average fewer running time than \emph{SUBSIM} while \emph{HyperIM} consumes at least $368X$ fewer time \emph{TriangleIM} with different sizes of the seed sets over the five hypergraphs. This is because when generating RR sets, the time complexity of \emph{HyperIM} is highly related to the number of layers of the vertices while that of \emph{SUBSIM} is highly related to the degrees of the vertices in the sample sets.
Since the number of layers is smaller that the degree of the vertices, \emph{HyperIM} performs fewer running time.
\emph{TriangleIM} uses the random edge sampling to generate triangle-based
RR sets without optimizations and it exhibits much more time than \emph{SUBSIM} and \emph{HyperIM}. Because \emph{HyperIM\_BRR} reduces the number of RR sets as explained in Section \ref{sec:optimization}, \emph{HyperIM\_BRR} can further reduce the running time to at least 1.56X. fewer than \emph{HyperIM}.
\subsection{Experimental results under the LT model}
We evaluate the four IM algorithms under LT model in terms of influence spread and the number of RR sets.
As shown in Figure \ref{fig_sample_SAMvaryIC}, when varying the error thresholds, \emph{HyperIM} and \emph{HyperIM\_BRR} always have higher influence spread than \emph{TriangleIM} and \emph{SUBSIM} as they can obtained more influenced vertices. It is noted that vary error thresholds almost do not cause the changes of  influence spreads because the four algorithms use the generations of RR sets to discover the seed sets whose approximation guarantees are solid. On the other hand, the numbers of required RR sets by \emph{HyperIM} and \emph{HyperIM\_BRR} are also smaller than \emph{TriangleIM} and \emph{SUBSIM} under the LT model as shown in Figure \ref{fig_sample_RR_LT}. The running times of the four algorithms are similar to those shown in Figure \ref{fig_sample_RR_LT}. These experimental results show that the stratified sampling combined with the two strategies is highly effective under both the IC and LT models.
\section{Conclusion}
In this paper, we propose \emph{HyperIM} to efficiently solve the IM problem in hypergraphs by stratified sampling to generate random reverse reachable sets. It is proven that \emph{HyperIM} is able to optimize the vertex selections and improve the accuracy while reducing the time complexity. Specifically, we carefully design the stratified sampling combined with a Binomial-based strategy and a Possion-based strategy to improve the efficiency and accuracy when generating the reverse reachable sets. We further propose HyperIM\_BRR by exploring the information of the stratified sampling to derive tighten bounds of the number of the RR sets to reduce the costs. The experimental results on the five hypergraph datasets confirm that our proposed algorithms outperform the state-of-the-art algorithms.
\section{Conclusion}
\bibliographystyle{abbrv}
\bibliography{Graph_ISO}
\end{document}